\documentclass[twocolumn,secnumarabic,amssymb, nobibnotes, aps, prb]{revtex4-2}
\usepackage{amsmath}
\usepackage{mathtools}

\usepackage{algorithmic}
\usepackage{graphicx}
\usepackage{textcomp}
\usepackage{xcolor}
\usepackage{caption}
\usepackage{subfigure}
\usepackage{booktabs} 
\usepackage{bm}
\usepackage{multirow}
\AtBeginDocument{ 
\heavyrulewidth=.08em
\lightrulewidth=.05em
\cmidrulewidth=.03em
\belowrulesep=.65ex
\belowbottomsep=0pt
\aboverulesep=.4ex
\abovetopsep=0pt
\cmidrulesep=\doublerulesep
\cmidrulekern=.5em
\defaultaddspace=.5em
}

\newcommand*{\mrm}[1]{\mathrm{#1}}
\newcommand*{\mtrm}[1]{\relax\ifmmode\mrm{#1}\else{#1}\fi} 
\renewcommand*{\vec}[1]{\bm{\mrm{#1}}}
\renewcommand*{\tensor}[1]{\bm{#1}}

\newcommand*{\CrCl}{$\rm{CrCl_3}$}
\newcommand*{\CrGeTe}{$\rm{CrGeTe_3}$}
\newcommand*{\CrBr}{$\rm{CrBr_3}$}
\newcommand*{\CrI}{$\rm{CrI_3}$}
\newcommand{\eg}[1]{{\textit{e.g.,} #1}}

\newcommand*{\Curie}{\mtrm{C}} 
\newcommand*{\TCurie}{\relax\ifmmode{\textit{T}_{\Curie}}\else{\textit{T}\textsubscript{\Curie}}\fi} 

\newcommand*{\nth}[1]{\relax\ifmmode{^{\text{#1}}}\else{\textsuperscript{#1}}\fi} 
\newcommand*{\Over}[1]{\frac{1}{#1}}

\newcommand*{\Vd}{$V_\mathrm{ds}$}
\newcommand*{\bcdot}{\mathbin{\bm{\cdot}}}
\DeclarePairedDelimiterX{\norm}[1]{\lVert}{\rVert}{#1}

\begin{document}
\title{Atomistic modeling of spin and electron dynamics in two-dimensional magnets switched by two-dimensional topological insulators}
\author{Sabyasachi Tiwari$^{1,2,3}$, Maarten~L.~Van~de~Put$^{1,3}$, Kristiaan Temst$^{4}$, William~G.~Vandenberghe$^{1}$, and Bart~Sor\'ee$^{3,}$$^{5,}$$^{6}$}
\address{$^1$ Department of Materials Science and Engineering, The University of Texas at Dallas, 800 W Campbell Rd., Richardson, Texas 75080, USA}
\address{$^2$Department of Materials Engineering, KU Leuven, Kasteelpark Arenberg 44, 3001 Leuven, Belgium}%
\address{$^3$ Imec, Kapeldreef 75, 3001 Heverlee, Belgium}%
\address{$^4$Quantum Solid State Physics, Department of Physics and Astronomy, KU Leuven, Celestijnenlaan 200 D, B-3001 Leuven, Belgium}%
\address{$^5$Department of Electrical Engineering, KU Leuven, Kasteelpark Arenberg 10, 3001 Leuven, Belgium}%
\address{$^6$Department of Physics, University of Antwerp, Groenenborgerlaan 171, 2020 Antwerp, Belgium.}


\begin{abstract}

To design fast memory devices, we need material combinations which can facilitate fast read and write operation.
We present a heterostructure comprising a two-dimensional (2D) magnet and a 2D topological insulator (TI) as a viable option for designing fast memory devices.
We theoretically model spin-charge dynamics between the 2D magnets and 2D TIs.
Using the adiabatic approximation, we combine the non-equilibrium Green's function method for spin-dependent electron transport, and time-quantified Monte-Carlo for simulating magnetization dynamics.
We show that it is possible to switch the magnetic domain of a ferromagnet using spin-torque from spin-polarized edge states of 2D TI.
We further show that the switching between TIs and 2D magnets is strongly dependent on the interface exchange ($J_{\mathrm{int}}$), and an optimal interface exchange depending on the  exchange interaction within the magnet is required for efficient switching.
Finally, we compare the experimentally grown Cr-compounds and show that Cr-compounds with higher anisotropy (such as \CrI)~results in lower switching speed but more stable magnetic order.

\end{abstract}
\maketitle
\section{Introduction}
Thanks to the recent discovery of two-dimensional (2D) magnets \eg \CrI~\cite{CrI_exp}, \CrBr~\cite{CrBr_exp}, and \CrGeTe~\cite{CrGeTe_exp}, research into 2D magnetics has garnered unprecedented attention.
2D magnetic materials open a plethora of opportunities in their use in future application in devices including spintronics~\cite{sample21,Nat_comm}, valleytronics~\cite{sample37}, and skyrmion~\cite{Nat_comm2}-based magnetic memories~\cite{sample22}.
However, many of the devices require low-dimensional magnets interfaced with semiconductors to function as memory devices ~\cite{TI-FM1, TI-FM2, TI-FM3}.

An interesting avenue of designing electronic devices using low-dimensional magnets lies in interfacing low-dimensional ferromagnets (FM) with topological insulators (TI)~\cite{TI-FM1,TI-FM2}.
The surface states of topological insulators can act as spin-channels with high spin-polarizability.
Moreover, depending on the direction of applied bias, the spin of the edge states can be switched thanks to spin-momentum locking~\cite{TI-FM4}.

Although there have been some experimental works on realizing magnetic devices using TI-FM interfaces~\cite{TI-FM1,TI-FM2, TI-FM3, TI-FM4}, a solid theoretical understanding of the interfacial physics is missing.
Most of theoretical works have either investigated equilibrium TI-FM interfaces with a FM having a fixed magnetic orientation~\cite{TI-FM5, TI-FM-hetero1} or investigated the impact of magnetic materials on the topological order of the TIs~\cite{TI-FM-hetero2}.

For the technological application of TI-FM material systems, it is necessary to understand how the motion of spin-polarized charge carriers in TIs impact the spin dynamics of the FM and vice-versa.
To understand such a coupled spin-charge effect, it is necessary to model the coupled spin-charge dynamics of TIs and FMs.
There have been recent theoretical works on modelling FM-semiconductor interfaces~\cite{Javier2020, Javier-JAP, Javier-3D-TI}, however a major issue lies with the description of the 2D magnets.

Most of the current methods use $\rm Landau–Lifshitz–Gilbert$ (LLG)~\cite{LLG} equation to describe magnetization dynamics, coupled with quantum transport methodologies, e.g., non-equilibrium Green's function (NEGF)~\cite{NEGF-LLG}.
The largest limitation lies in the magnetization dynamics of the magnetic material because the LLG equation assumes a continuous description of the magnetic structure~\cite{Javier2020} and the atomistic description is lost.
Moreover, interactions such as exchange anisotropy, which play a major role in determining the magnetic order of low-dimensional magnets~\cite{Tiwari2021_Curie, my-paper, Vanherck18, Vanherck20, Tiwari2021} are hard to include in the LLG equation, and most frequently a scalar approximation to the exchange interactions is assumed~\cite{LLG-bad}.

On the other hand, Monte-Carlo (MC) simulations are shown to be very accurate in predicting temperature dependent observables for 2D magnets taking into account full anisotropy~\cite{my-paper, Tiwari2021_Curie, Lado_2017}.
Unfortunately, MC simulations do not have a standard method of quantifying time to obtain time dependent observables such as magnetic switching.
There have been previous works on quantifying time within MC simulations for systems with in-plane rotational invariance~\cite{TQMC1, TQMC2}, called time-quantified Monte-Carlo (TQMC) simulations.
Thankfully, most of the 2D magnetic materials yet discovered experimentally have an in-plane rotational invariance, and TQMC can be applied for such materials for magnetization dynamics. 

\begin{figure*}[htp]
   \centering
    {\includegraphics[width=1.7\columnwidth]{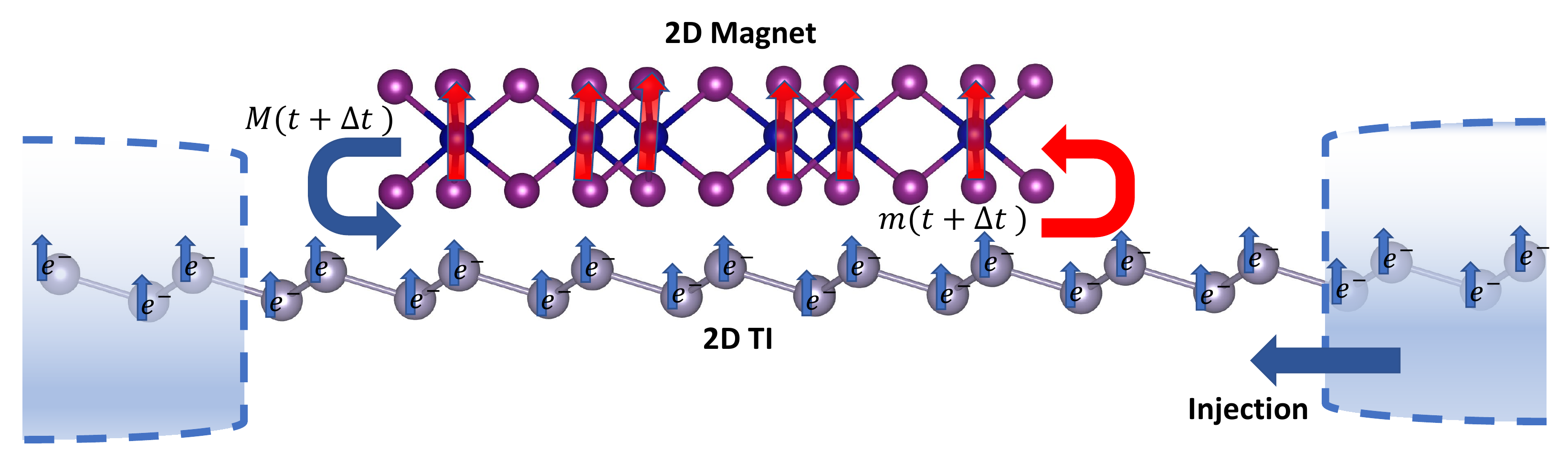}}   
    \caption{The interface between a 2D TI and a layered magnet. The electrons are injected from the contacts (illustrated by the blue arrow below right contact) at time $t+\Delta t$. The flowing electrons carry a small magnetic moment (illustrated by the small blue arrows) exerting a torque due to electronic magnetic configuration: $m(t+\Delta t)$ on the interfacing magnet with magnetic configuration: $M(t+\Delta t)$ (illustrated by the red arrows).}
	\label{f:TQMC}
\end{figure*} 

We present a theoretical study of spin-dynamics and spin-induced switching in a TI-FM heterostructure.
We combine the non-equilibrium Green's function method for spin-dependent electron transport with time-quantified Monte-Carlo for simulating magnetic systems.
We use our method to study spin-induced switching in a heterostructure of a two-dimensional topological insulator and a two-dimensional ferromagnet.
We show that it is possible to change the magnetic-domain structure in the ferromagnet using spin-injection from TIs, which can be used to design high-speed memory devices.
We then show that the switching can only be achieved efficiently for an optimal interfacial exchange interaction between TI and FM.
Finally, we compare the switching time for four experimentally grown Cr-compounds: \CrI, \CrBr, \CrCl, and \CrGeTe. 
We show that the higher anisotropy of \CrI~results in a much larger switching time compared to \CrBr~and \CrCl, which have lower anisotropy.
 
\section{Methodology}\label{s:method}

\subsection{The Heisenberg Hamiltonian}
We model the 2D magnetic structure using the Heisenberg Hamiltonian~\cite{Tiwari2021_Curie}
\begin{equation}
	H_{\mathrm{m}}
	=
	-\Over{2}\sum_{i\neq j}\vec{{S}}_i \cdot \tensor{J}_{ij} \cdot \vec{{S}}_j
	+D \sum_{i} ({S}_i^z)^2
\label{e:Heisenberg1}
\end{equation}
where, $\vec{{S}}={S}^{\vec{x}} \vec{x}+{S}^{\vec{y}} \vec{y}+{S}^{\vec{z}} \vec{z}$ is the spin-vector with magnitude $|\mathbf{S}|=\sqrt{(S^\mathbf{x})^2+(S^\mathbf{y})^2+(S^\mathbf{z})^2}$.   
The exchange interaction strength  $\tensor{J}_{ij}$ between spins at site $i$ and $j$ is a $3\times 3$ tensor~\cite{my-paper} whose parameters are obtained by fitting to DFT calculations~\cite{my-paper, Tiwari2021_Curie}.
The second term in Eq.~(\ref{e:Heisenberg1}) is the single-ion anisotropy with strength $D$. 
We assume in-plane rotational invariance for the 2D magnetic materials which leads to the Hamiltonian reducing to,
\begin{multline}
H_{\mathrm{m}} = \sum_{i,j} \frac{J_{ij}}{2} \left[\vec{{S}}_i \bcdot \vec{{S}}_j + \Delta_{ij} ({S}^{\vec{z}}_i {S}^{\vec{z}}_j - {S}^{\vec{x}}_i {S}^{\vec{x}}_j - {S}^{\vec{y}}_i {S}^{\vec{y}}_j)\right]\\	 
	+D \sum_{i} ({S}_i^{\vec{z}})^2,
	\label{e:Heisenberg}
\end{multline}
where the exchange anisotropy ($\Delta_{ij}$) accounts for the distinct values for the in-plane and out-of-plane anisotropic exchange strength, $J_{ij}^{\vec{xx}}=J_{ij}^{\vec{yy}}=J_{ij}(1-{\Delta_{ij}})$ and $J_{ij}^{\vec{zz}}=J_{ij}(1+{\Delta_{ij}})$~\cite{Tiwari2021_Curie}, respectively.

Moreover, we define the total magnetization per atom in the $\vec{x}$, $\vec{y}$, and $\vec{z}$ direction as,
\begin{equation}
	S_{\vec{{z/y/x}}}=\frac{1}{N_\mathrm{atom}}\sum_i S^{{\vec{z/y/x}}}_i.
\end{equation}
Here, $N_{\rm atom}$ is the total number of atoms.

\subsection{Electronic Hamiltonian}
We model the electronic structure of the topological insulators using a tight-binding Hamiltonian,
\begin{multline}
	H_{\mathrm{elec}}
	=	-t\sum_{\langle r,r' \rangle,\alpha} c_{r,\alpha}^{\dagger} c_{r',\alpha}\\
    + \mathrm{i} \Lambda_\mathrm{so} \sum_{\langle \langle r,r' \rangle \rangle,\alpha,\beta} 
      v_{r,r'} c_{r,\alpha}^\dagger \sigma_{\alpha,\beta}^{\vec{z}} c_{r',\beta}.
\label{e:TB}
\end{multline}
Here the first term, with coupling strength $t$, accounts for the nearest neighbor hopping, $\langle r,r'\rangle$, between adjacent lattice sites $r$ and $r'$, with respective electron creation and annihilation operators $c_{r,\alpha}^{\dagger},\,c_{r',\alpha}$. 
$\alpha$ and $\beta$ represent spin degrees of freedom, \textit{i.e.}, $\alpha \in {\{\uparrow, \downarrow\}}$ and $\beta \in {\{\uparrow, \downarrow\}}$.
The second term is the next nearest neighbor spin-orbit coupling term with strength $\Lambda_\mathrm{so}$. 
$\sigma^{\vec{z}}$ is the $\rm{z}^{th}$ component of the Pauli matrices.
The parameter $v_{r,r'}$ is $+1$ when the shortest next nearest neighbor path from $r$ to $r'$ with respect to atom $r$ is clockwise, and $-1$ if it is counterclockwise.

\subsection{Combined Hamiltonian}

We combine the model for the Heisenberg Hamiltonian and the electronic Hamiltonian using the interaction Hamiltonians 
\begin{subequations}
\begin{align}
	H'_{\mathrm{m}} & =J_{\mathrm{int}}\sum_{r,i}\vec{m}_r \cdot \vec{S}_i, \\
	H'_{\mathrm{elec}} &=J^{\mathrm{e}}_{\mathrm{int}}\sum_{i,r,\alpha,\beta}c_{r,\alpha}^{\dagger}\vec{S}_i \cdot \vec{\sigma}_{\alpha,\beta}c_{r,\beta}.
\end{align}
\end{subequations}
Here, the $J_{\mathrm{int}}$ and $J^{\mathrm{e}}_{\mathrm{int}}$ are the interface interactions at the semiconductor-FM interface.
The electron magnetization, $\vec{m}={m}^{\vec{x}}\vec{x}+{m}^{\vec{y}}\vec{y}+{m}^{\vec{z}}\vec{z}$ is the expectation value of the Pauli spin matrix: $\vec{\sigma}=\sigma^{\vec{x}}\vec{x}+\sigma^{\vec{y}}\vec{y}+\sigma^{\vec{z}}\vec{z}$.
Explicitly, we calculate the electron magnetization $\vec{m}$ by taking the trace of the density matrix over the spin degrees of freedom, 
\begin{equation}
	\vec{m} (r) =\frac{\hbar\gamma}{4\pi}\mathrm{Tr}[\vec{\sigma} \cdot \rho (r,r')]_{\mathrm{spin}},
\end{equation}
with the electronic density matrix,
\begin{equation}
	\rho (r,r') =\int{G^{<}(r,r',E) dE},
\end{equation}
where, $G^{<}(E)$ is the lesser Green's function.
As is standard in ballistic NEGF, the lesser Green's function is obtained from,~$G^<(r,r',E) = A_\mathrm{L} (r,r',E) f(E-\mu_\mathrm{L})+ A_\mathrm{R} (r,r',E) f(E-\mu_\mathrm{R})$.
Here, $A_{\mathrm{L}/\mathrm{R}}(r,r',E)$ are the spectral functions for the left ($\rm L$) and the right ($\rm R$) contacts, and $\mu_{\rm L/R}$ are the respective chemical potentials with $f(E)$ being the Fermi-Dirac distribution function.
The spectral functions are obtained using, $A_\mathrm{L/R} = G^\mathrm{r} \Gamma_\mathrm{L/R} G^\mathrm{a}$, with $\Gamma_{\mathrm{L/R}}=i(\Sigma_{\mathrm{L/R}}-~\Sigma^{\dagger}_{\mathrm{L/R}})$.
$G^\mathrm{r}(E) =(E-H'_{\mathrm{elec}}-\Sigma_{\mathrm{L}}-\Sigma_{\mathrm{R}})^{-1}$ is the retarded Green's function.
The advanced Green's function is $G^\mathrm{a}=(G^\mathrm{r})^{\dagger}$.
The contact self-energies $\Sigma_{\mathrm{L/R}}$ are obtained using the quantum transmitting boundary method (QTBM)~\cite{QTBM, Tiwari_2019}.
Similar to magnet, we define the magnetization per atom in TIs as,
\begin{equation}
	m_{\vec{{z/y/x}}}=\frac{1}{N^{\mathrm{TI}}_\mathrm{atom}}\sum_i m^{\vec{z/y/x}}_i.
\end{equation}
Here, $N^{\mathrm{TI}}_{\rm atom}$ is the total number of atoms in TI.

\begin{figure}[htp]
	{\includegraphics[width=0.6\columnwidth]{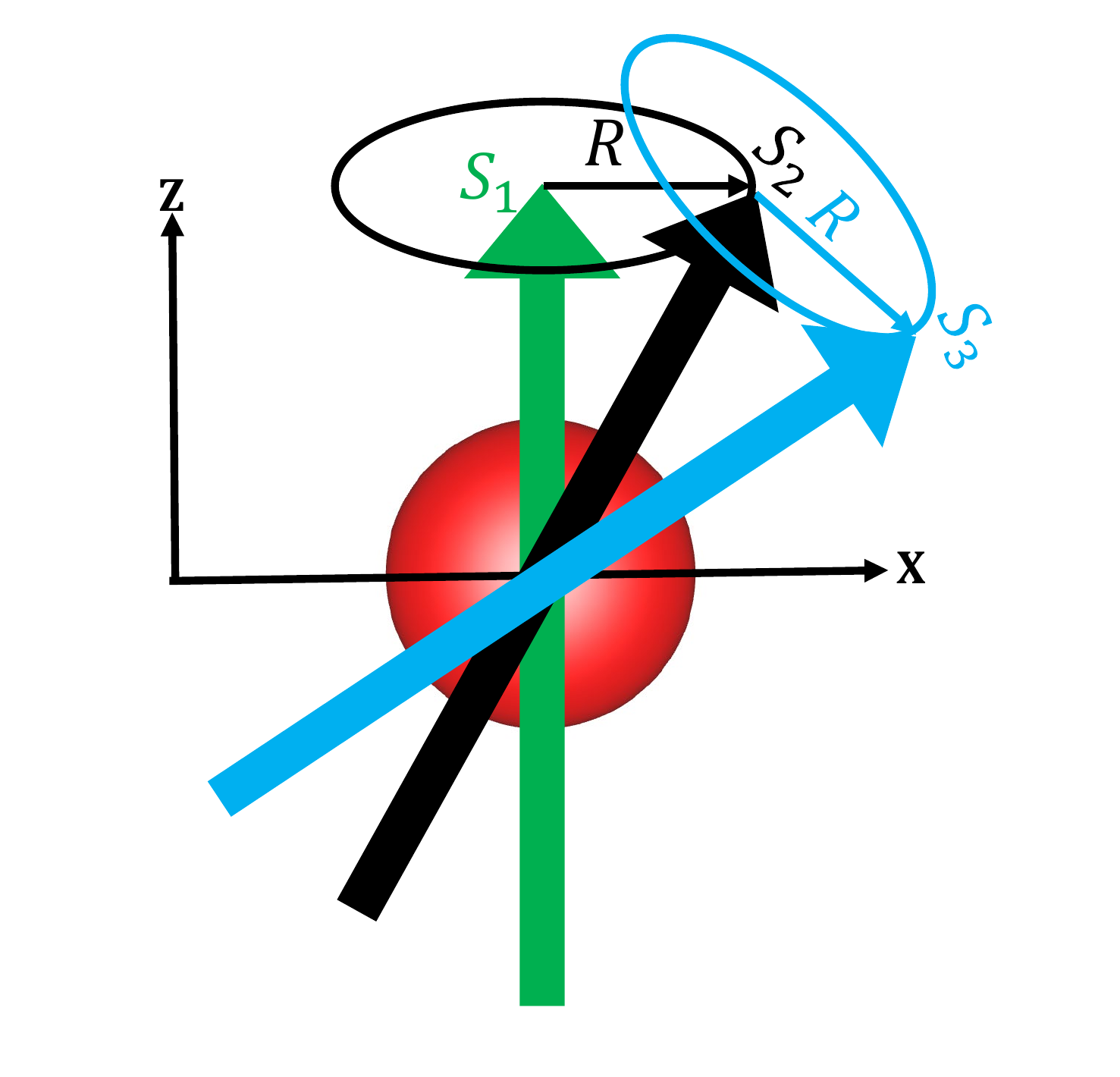}}
	\caption{The choice of spin in a MC step using TQMC. The Spin $S_1$ shows the initial spin state which is allowed to take a value within the radius $R$ with spin state ranging between $S_1$ to $S_2$ (illustrated in black circle). The spin state $S_2$ shows an intermediate spin state which can take a value within the the radius $R$ (illustrated in blue) with spin state switching from $S_2$ to $S_3$. The radius $R$ is determined by the Gilbert damping.}
\label{f:TQMC2}
\end{figure}

\begin{figure*}[htp]
  \centering  
  \subfigure[]{\includegraphics[width=\columnwidth]{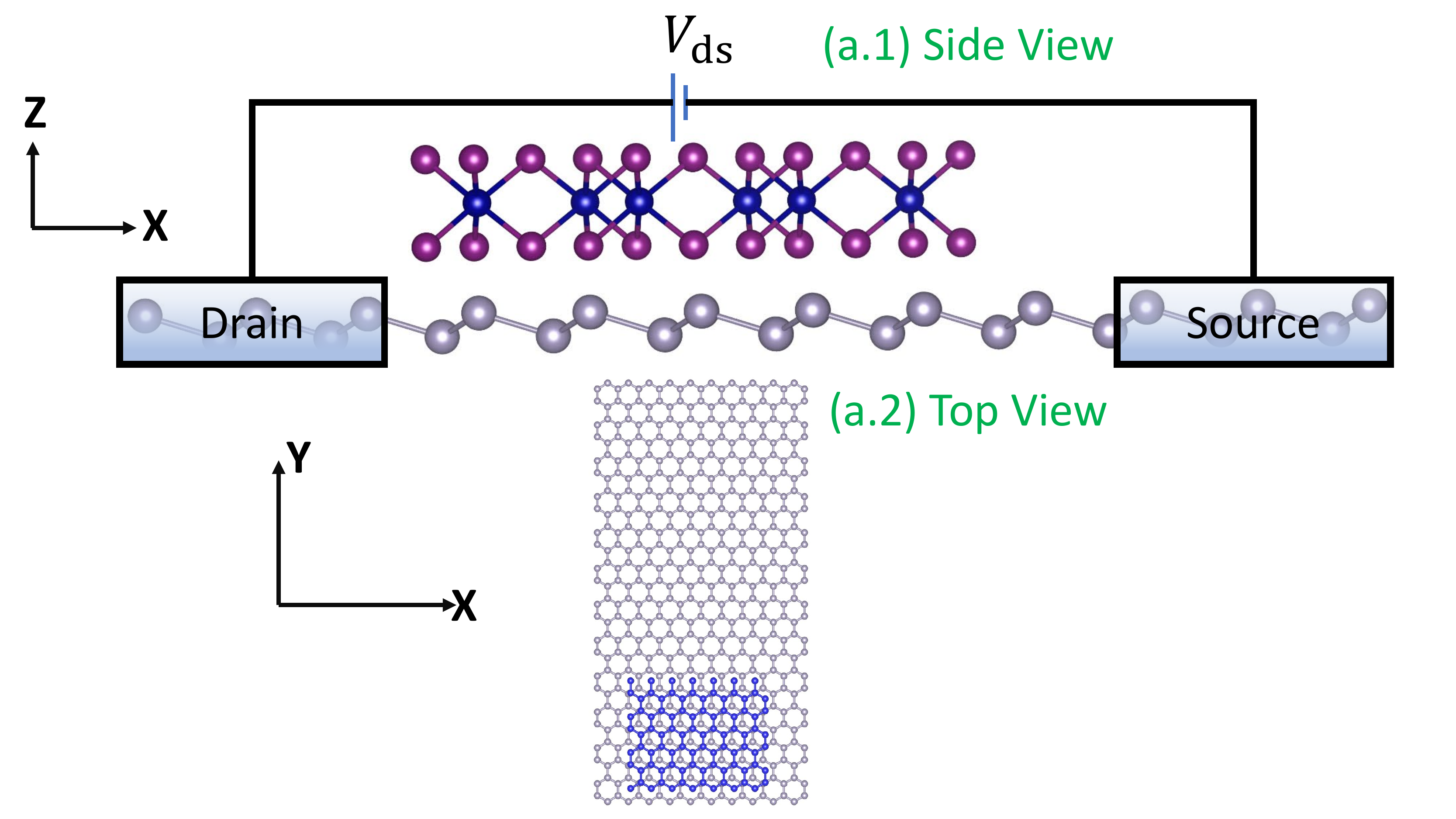}}
  \subfigure[]{\includegraphics[width=0.75\columnwidth]{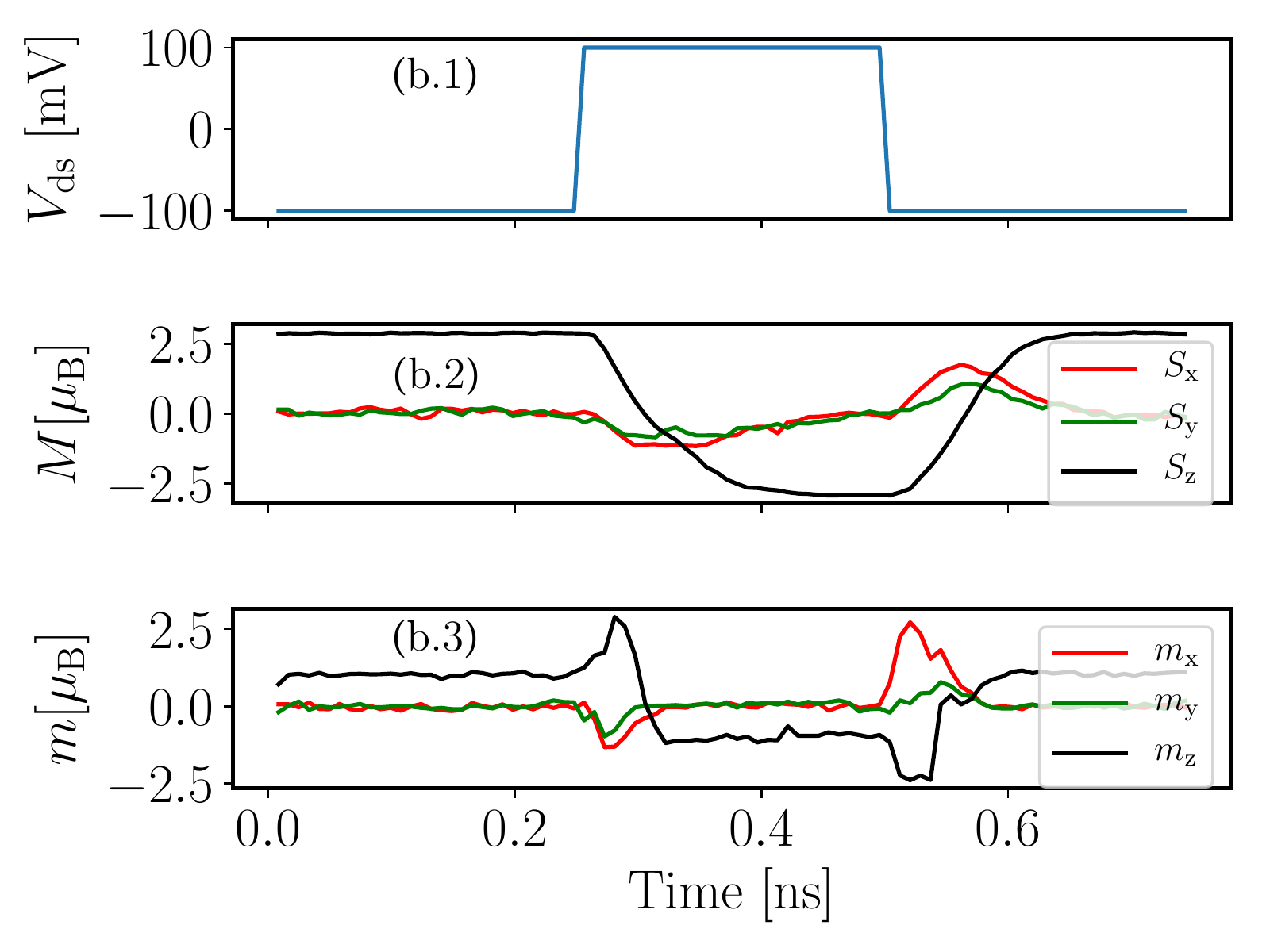}}
  \caption{(a) The device configuration studied for TI-FM switching. The 2D TI is contacted with a drain and a source contact and a voltage ($V_{\rm ds}$) is applied between the contacts. (b) The magnetization of the 2D FM as a function of time. (b.1) The bias is applied at $t=0\, \rm ns$ and switched to 100 meV from -100 meV at 0.25 ns. (b.2) The magnetization of the 2D FM in the $\vec{z}$-direction ($S_{\rm z}$) switches its direction (from $3\, \mu_{\rm B}$ to $-3\, \mu_{\rm B}$). (b.3) The induced magnetization in the 2D TI.} 
  \label{f:switch_device} 
\end{figure*} 

\begin{figure}[htp]
	\centering   
    {\includegraphics[width=1\columnwidth]{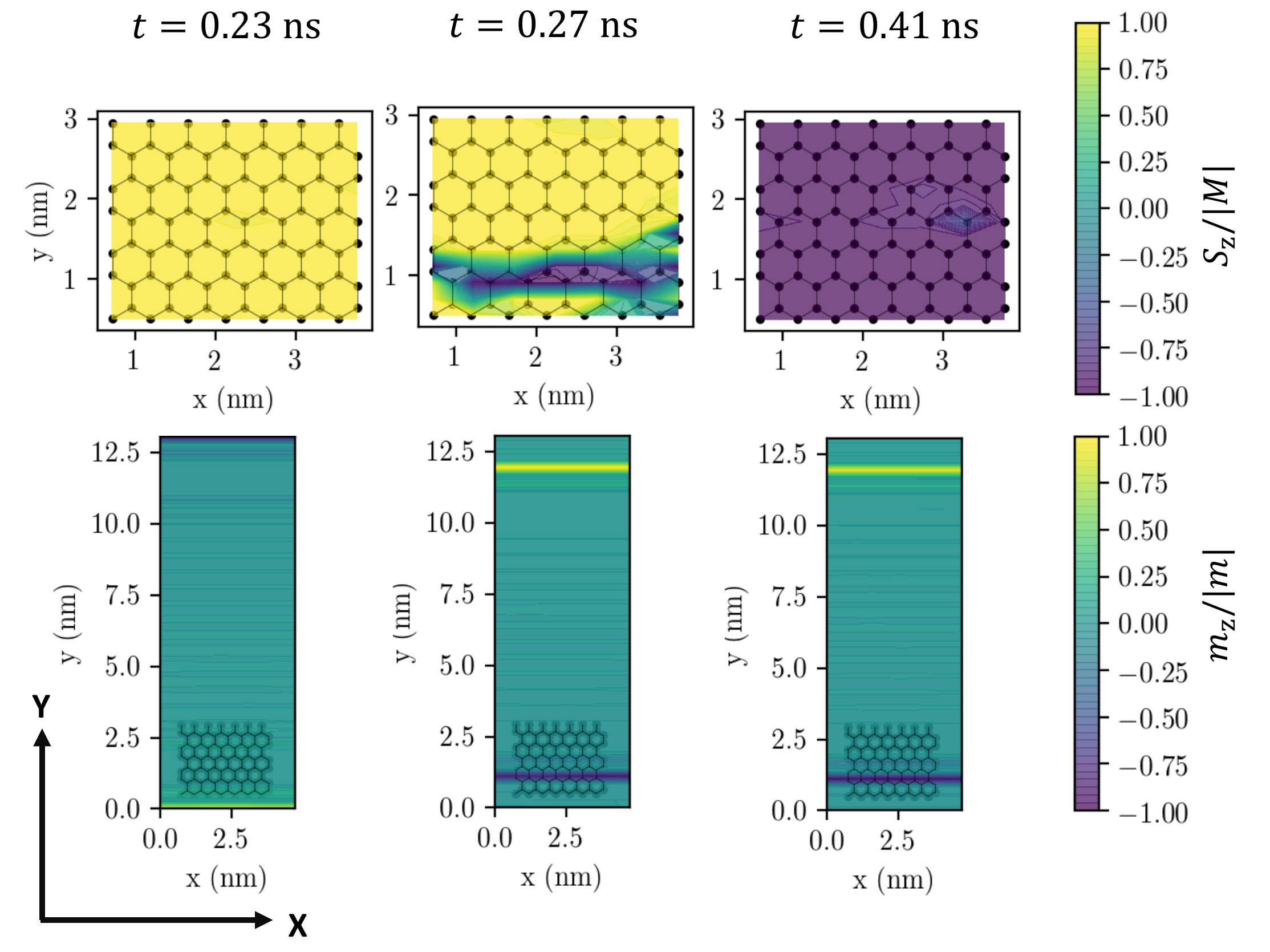}} 
    \caption{The magnetic configuration ($M(t)$) of the FM (upper three panels), and the electronic spin configurations ($m(t)$) of the TI (lower three panels) at $t=0.23 \rm ns,\,0.27$ ns, and $0.41$ ns while the voltage pulse is switched at $t=0.25$ ns. The atomic structure of the FM is shown as an imprint within the magnetic configuration of the TI.}
	\label{f:magnetic_domain}
\end{figure} 

\begin{figure*}[htp]
	\centering   
   \subfigure[]{\includegraphics[width=0.9\columnwidth]{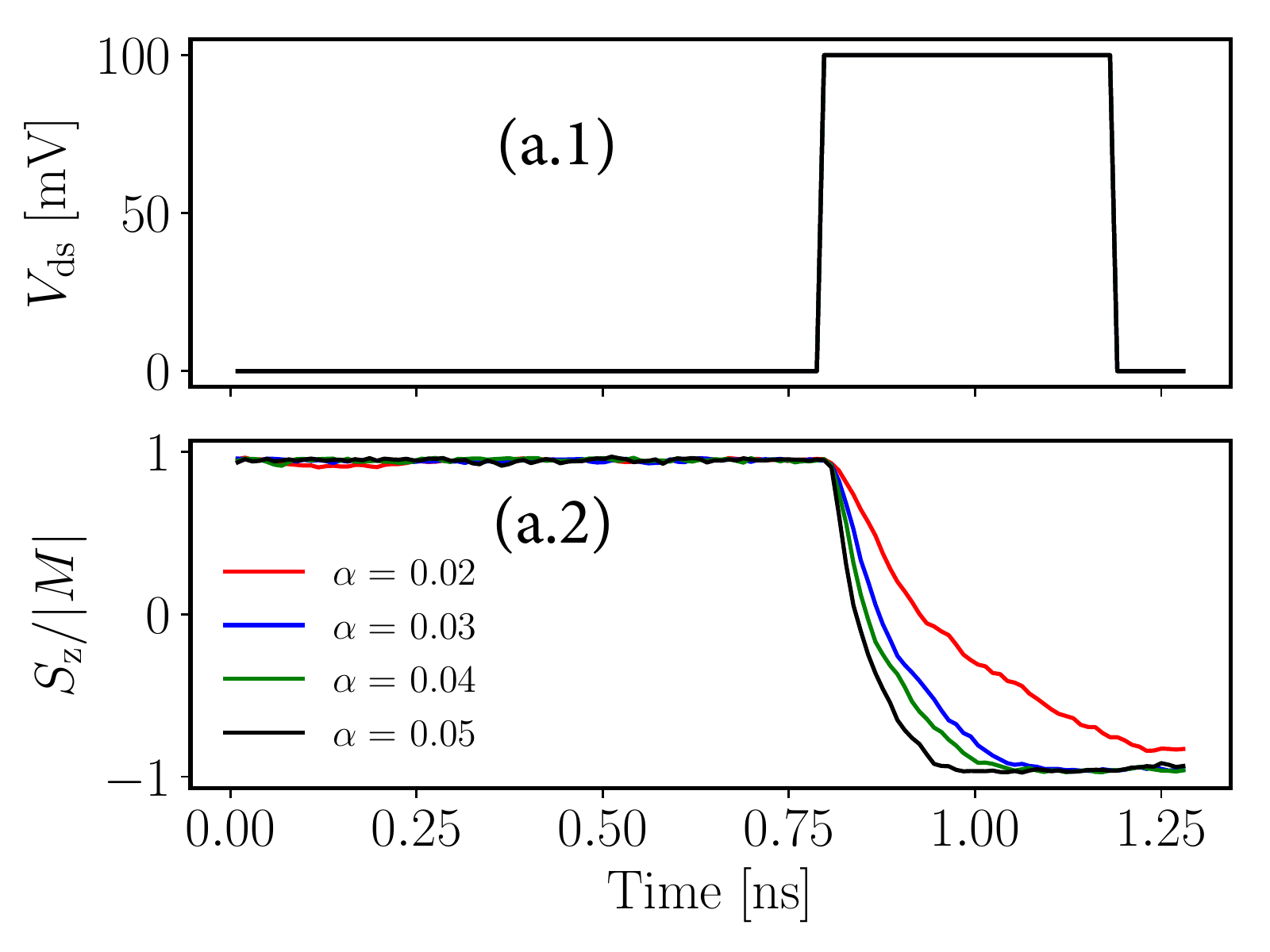}} 
   \subfigure[]{\includegraphics[width=0.9\columnwidth]{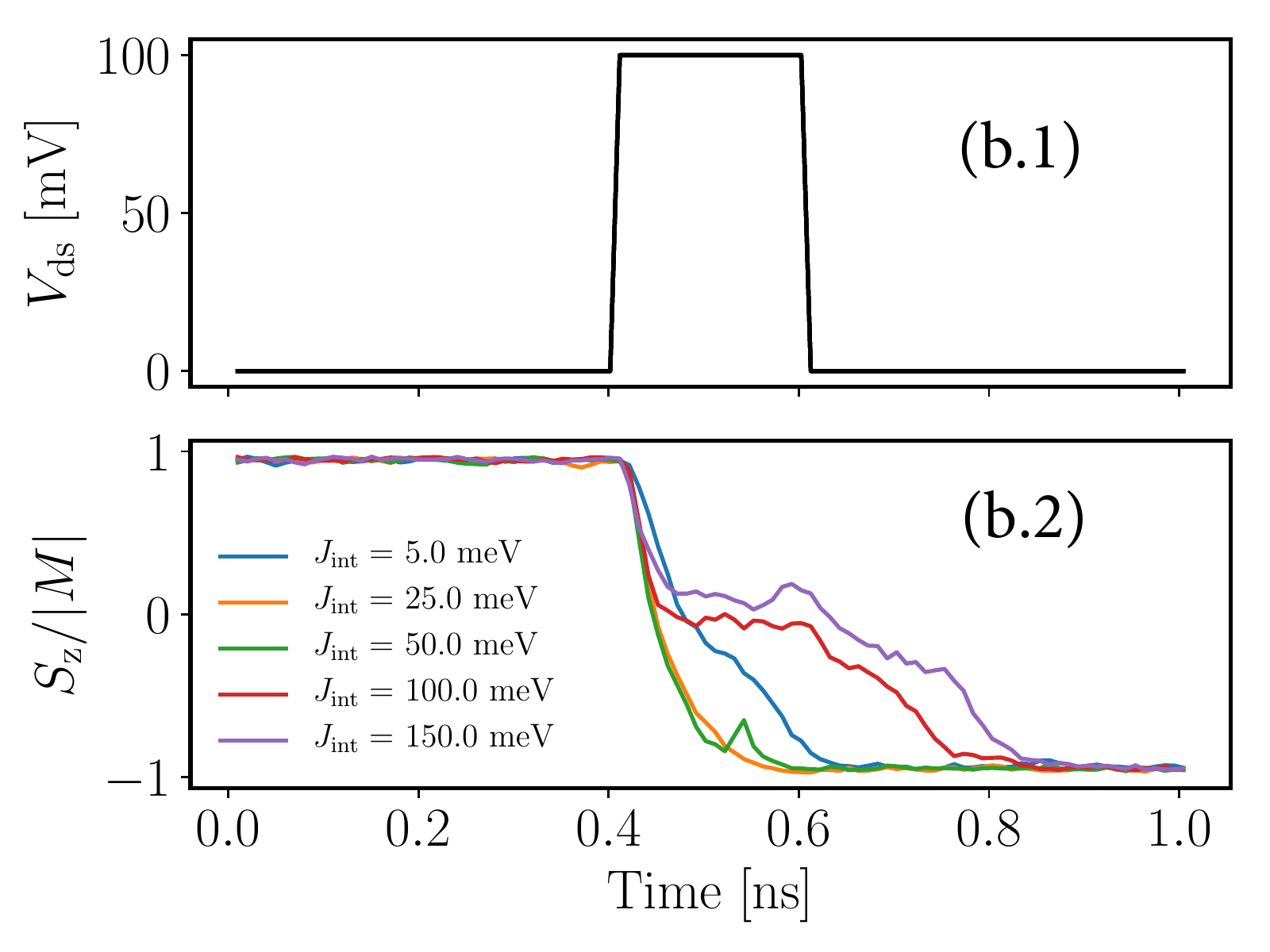}}\\ 
   \subfigure[]{\includegraphics[width=0.9\columnwidth]{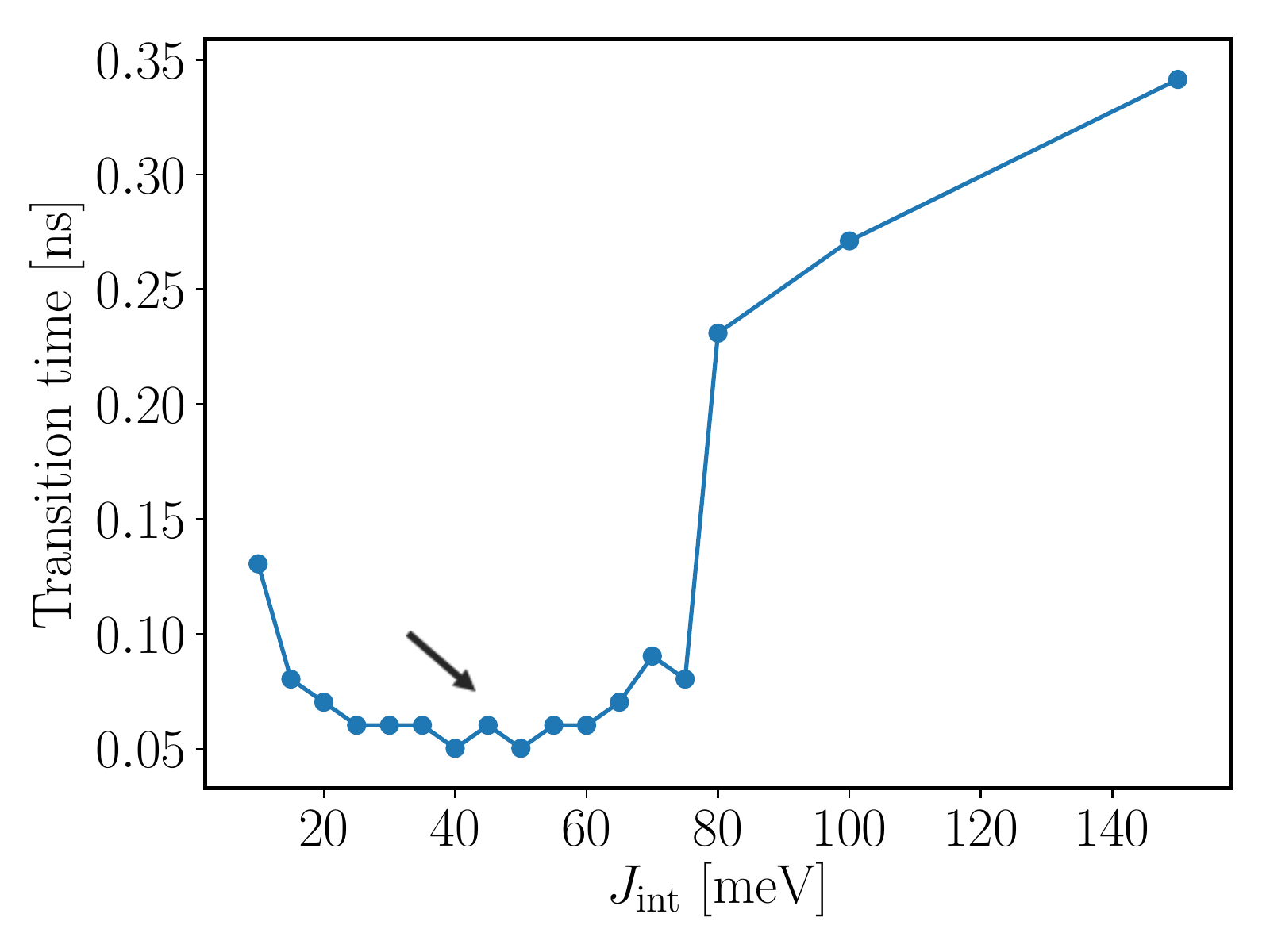}} 
      \subfigure[]{\includegraphics[width=0.7\columnwidth]{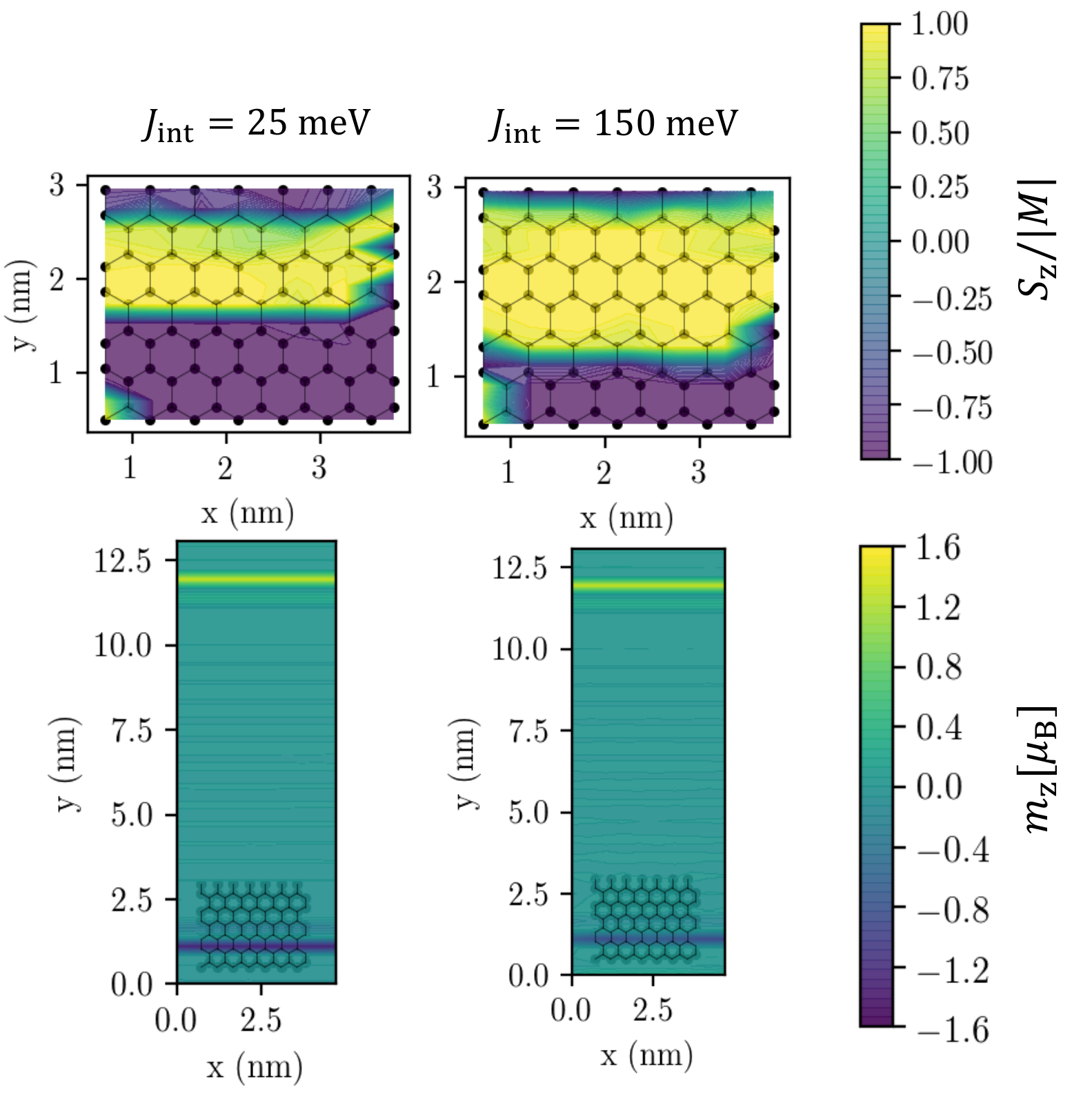}} 
    \caption{(a) The applied \Vd~(panel a.1) and the magnetization of the top 2D FM for $\alpha=\,0.02,\,0.03,\,0.04,\,0.05$ (panel a.2). (b)~The applied \Vd~and the magnetization of the top 2D FM for interface interaction strengths $J_{\mathrm{int}}=\,5,\,25,\,50,\,100,\,150$ meV (panel b.2). (c) The transition time for the applied pulse in panel (b.1) as a function of $J_{\rm int}$. (d) The magnetization at $t=0.45$~ns for $J_{\rm int}=25$~meV and $J_{\rm int}=150$~meV. The upper two panels show the magnetization of the 2D FM in the z-direction ($S_{\rm {z}}/|M|$), and the lower two panels show the induced magnetization in the 2D TI ($m_{\rm {z}}$).}
	\label{f:damping}
\end{figure*} 
\subsection{Magnetization dynamics and time-quantification}
To simulate the magnetization dynamics as a function of time, we use the time-quantified Monte-Carlo (MC) approach~\cite{TQMC1, TQMC2}.
Within TQMC, one step of a MC simulation is assigned a time step $\Delta t$ of,
\begin{equation}
	R^2=\frac{20 k_{\mathrm{B}}T\alpha\gamma}{(1+\alpha^2)M}\Delta t.
\label{e:TQMC}
\end{equation}
Here, $R$ is a cone radius up to which a trial spin-rotation is allowed in each MC step. 
$k_{\mathrm B}$ is the Boltzmann constant, $\alpha$ is the Gilbert damping parameter, $M=\sqrt{S^2_\mathbf{x}+S^2_\mathbf{y}+S^2_\mathbf{z}}$ is the magnetic moment, and $\gamma$ is the gyromagnetic ratio.

Figure~\ref{f:TQMC2} shows a single magnetic atom with spins chosen for three successive time steps using TQMC ($S_1$, $S_2$, $S_3$).
Within TQMC for each time step, we make a cone of radius of $R$ around the spin-vector of the present spin configuration ($S_1$, green). 
The next spin-configuration ($S_2$, black) is chosen within the cone using the Metropolis algorithm~\cite{sample9, my-paper}.
Once, the spin-configration is updated to $S_2$, we choose the next spin configuration ($S_3$, blue) by using the same method of making a cone of radius $R$ around $S_2$.

The expression in Eq.~(\ref{e:TQMC}) for time-discretization has been obtained by assuming only single-spin interactions and Langevin dynamics~\cite{TQMC1, TQMC2}.
Therefore, this time discretization is valid only under the condition that the magnetization dynamics can be approximated using Langevin dynamics, and is more accurate for materials with in-plane rotational invariance and Gilbert damping: $\alpha \leq 1$~\cite{TQMC1, TQMC2}.
The approximation of magnetization dynamics to Langevin dynamics is only valid under high out-of-plane anisotropy or very low temperatures ($T<\,10$ K) where the spins start behaving collectively as a single unit.
Due to the above reason, we perform all our calculations below $T<\,10$ K.

\subsection{Algorithm for spin-charge dynamics}

The energy of the magnetic interaction is of the order of meV, while the electronic interaction energy is of the order of eV.
Therefore, the time scales of magnetization dynamics are in ps, whereas that of electrons is in fs.
Within the adiabatic approximation, we assume that due to the large differences in the time scale, the electrons are considered to only see an adiabatic change in magnetic moment and are thus always found to occupy  instantaneous eigenstates.

Hence, as shown in Fig.~\ref{f:TQMC}, we start with a magnetic orientation of the magnetic material, and at each time-step ($t+\Delta t$), we rotate the spin of the top magnetic layer using MC sampling.
After the MC step, we calculate the magnetic moment of the magnet $M(t+\Delta t)$.
For the same time-step, we obtain the lesser Green's function using the NEGF method and calculate the electronic magnetization $\mathbf{m}(t+\Delta t)$.
For the next time step, the magnetization of the electronic system is an input to the magnetic system, and the loop continues.

\subsection{Computational Details}
For all calculations unless mentioned specifically, we have used values $\alpha=0.05$, $k_{\mathrm{B}}T=0.5$ meV,
$J_{\mathrm{int}}=J^{\mathrm{e}}_{\mathrm{int}}=25\, \rm meV$.
We have used the parameters of a \CrI~monolayer to parameterize the interface 2D ferromagnet,
which we obtained from DFT calculations using the procedure in Ref.~\onlinecite{Tiwari2021_Curie}.
The saturation magnetization used for the Cr-atoms is $3\mu_{\rm B}$.
To model the TIs we have used the parameters of the tight-binding Hamiltonian for stanene~\cite{Tiwari_2019}.

\begin{figure*}[htp]
	\centering   
    \subfigure[]{\includegraphics[width=0.9\columnwidth]{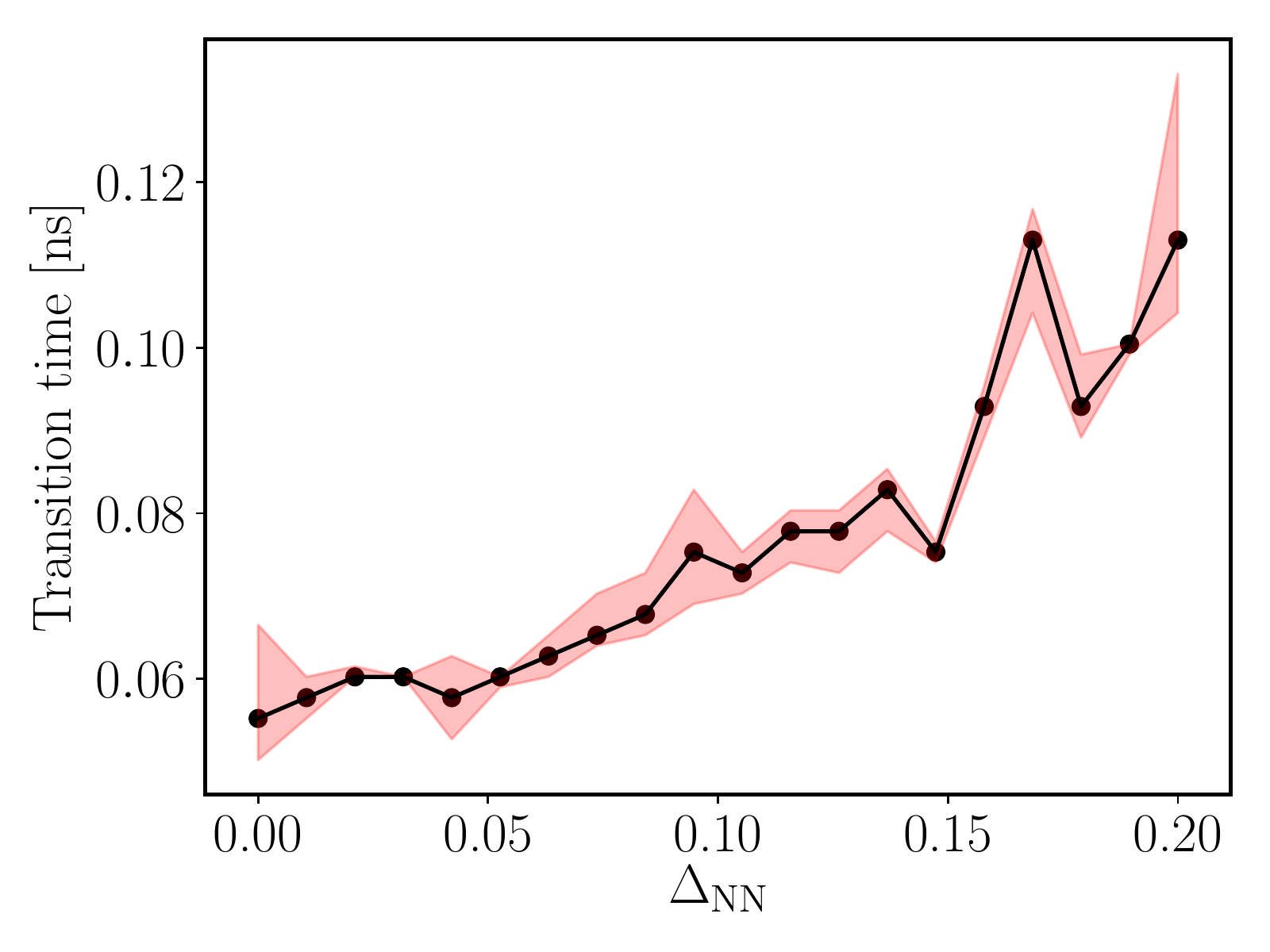}}
    \subfigure[]{\includegraphics[width=0.9\columnwidth]{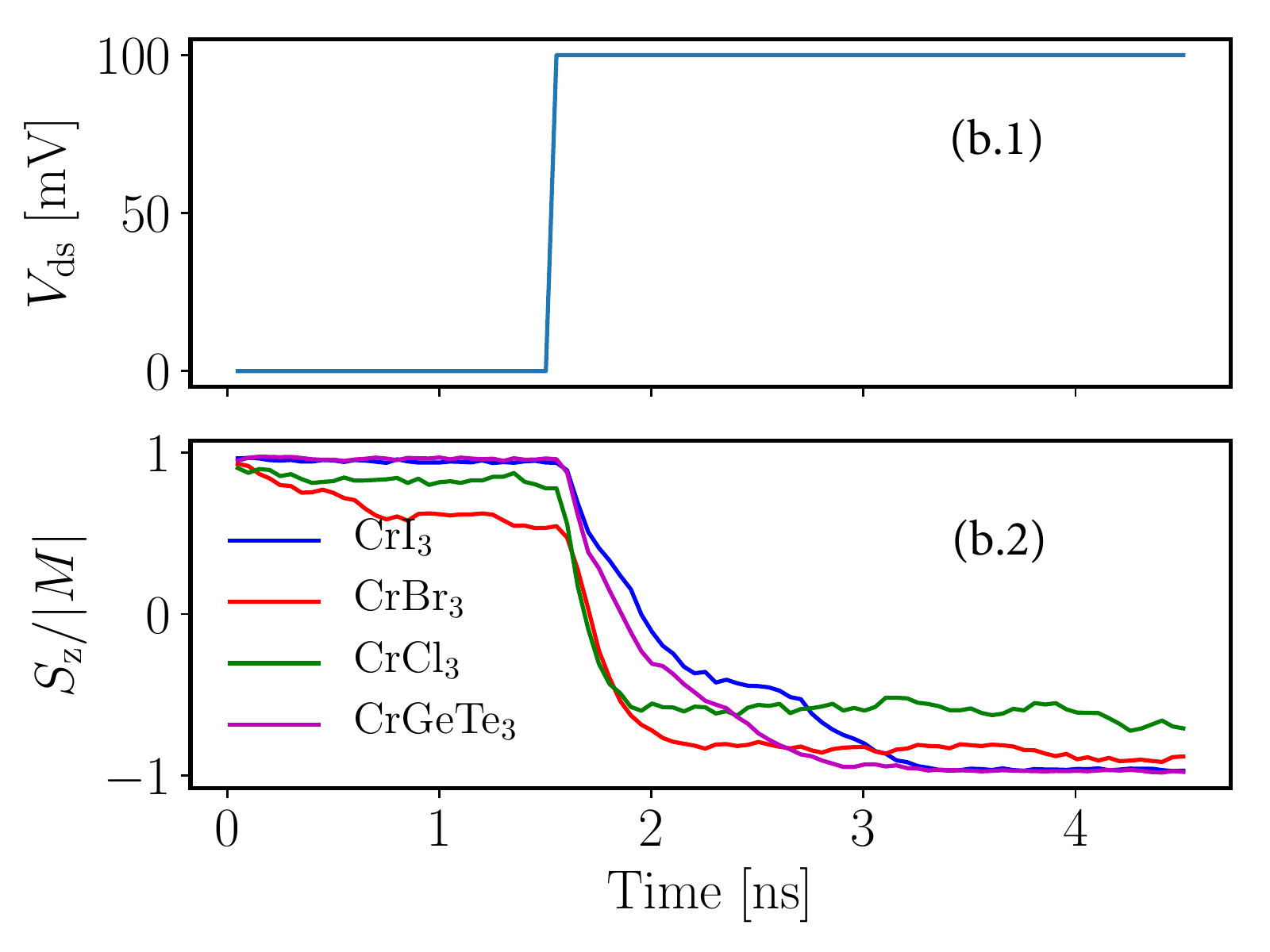}} 

     \caption{(a) Transition time as a function of nearest-neighbor anisotropy $\Delta_{\rm NN}$. We perform the same sweep of $\Delta_{\rm NN}$ for 10 different starting configurations. The solid line shows the median and the shaded region shows the $25^{\rm th}$-$75^{\rm th}$ percentile. (b) Comparison of switching in \CrI, \CrBr, \CrCl, and \CrGeTe. Panel (b.1) shows the applied \Vd~ and panel (b.2) shows the magnetization ($S_{\rm {z}}/|M|$) for \CrI, \CrBr, \CrCl, and \CrGeTe. We use an interface exchange $J_{\rm int}=25$~meV, and $\alpha=0.005$. }
	\label{f:Cr-compounds}
\end{figure*} 
\section{Results and Discussion}\label{s:results}

\subsection{TI-FM interface operation}

Figure~\ref{f:switch_device} (a) shows the TI-FM configuration we are investigating.
For the TI we use a 5 nm wide stanene ribbon, for the 2D FM we use \CrI~with size: $3\,\mathrm{nm}\times3.5\,\mathrm{nm}$.
The 2D FM is positioned near one of the edges of the 2D TI as shown in the top view in Fig.~\ref{f:switch_device} panel (a.2).
A potential ($V_{\mathrm{ds}}$) is applied across the TI ribbon to inject charge carriers in the TI ribbon.
For $V_{\rm ds}>0$, the edge states on the top edge of the TI have up-spin, whereas the bottom edge exhibit down-spin.
For $V_{\rm ds}<0$, spin polarization of the edge states is flipped.
Therefore, the interfaced FM experiences positive or negative spin-torque from the underlying charge carriers of the TI, depending on the applied $V_{\rm ds}$.

Figure~\ref{f:switch_device} (b) panel (b.1) shows the applied $V_{\rm ds}$ as a function of time.
We apply a step \Vd~pulse of 100 mV at $t=1.75$ ns.
Figure~\ref{f:switch_device} (b) panel (b.2) shows the average magnetization in the $\vec{x},\,\vec{y}$ and $\vec{z}$-direction of the top FM layer as a function of applied \Vd.
We observe that with a change in the applied \Vd~from -100 mV to 100 mV at $t=0.5$ ns, the magnetization in the z-direction ($S_{\mathrm{z}}$) switches from 3 $\mu_{\rm B}$ to -3 $\mu_{\rm B}$.
When the applied voltage switches from 100 mV to -100 mV, we observe that the magnetization switches again from -3 $\mu_{\rm B}$ to 3 $\mu_{\rm B}$.

Figure~\ref{f:switch_device} (b) panel (b.3) shows the induced magnetization of the TI ribbon in the $\vec{x}$, $\vec{y}$, and $\vec{z}$ direction as a function of time.
We observe that due to the magnetization of the top FM, there is a finite induced magnetization in the TI ribbons when the applied bias is non-zero.
Interestingly, we find that whenever the bias switches and the magnetization of the top FM transitions, the total induced magnetization of the TI ribbon also shows a peak.

To further analyze the magnetization dynamics of the TI-FM interface, we plot the out-of-plane projection of the magnetization of the entire TI and FM sample at various time steps in Fig.~\ref{f:magnetic_domain}.
Figure~\ref{f:magnetic_domain} shows the magnetization in the $\vec{z}$-direction for the 2D magnet (top panels, $S_{\rm {z}}/|M|$) and the TI (bottom panels, $m_{\rm {z}}/|M|$), respectively.
We observe that at $t=0.23 \, \rm ns$, the TI has a positive bias \Vd$>0$ and both the TI and the FM have positive magnetization at the interface.
Note that we have taken the interface parameter to be positive, meaning the interaction is ferromagnetic.

At $t=0.27 \, \rm ns$, the bias of the TI has opposite polarization \Vd $<0$.
We observe that the magnetization of the TI reverses.
The reversal of the magnetization in TI induces a spin-torque on the interfacing FM, which results in the formation of a small region of magnetization, with opposite orientation as that of the entire FM ($\hat{S}_{\mathrm{z}}<0$).

At $t=0.41 \, \rm ns$, we observe that the domain with $\hat{S}_{\mathrm{z}}<0$ grows and changes completely to a FM configuration with a magnetization of the 2D magnet equal to $-3 \mu_{\rm B}$.
Interestingly, we find that the top magnetization of the FM does not have a significant impact on the magnetization of the TI.
The reason for such behavior is because we have assumed in our calculations that the interface interaction strength $J_{\mathrm{int}}=J^{\mathrm{e}}_{\mathrm{int}}=25\, \rm meV$.
The interaction strength felt by the TI is of the order of meV, whereas the topological energy gap of TIs is of the order of eV.
Hence, the weak interaction strength will always lead to a negligible response of the TI magnetization due to magnetization at the interface with the FM.

\subsection{Impact of $J_{\rm int}$ and $\alpha$}

Figure~\ref{f:damping} (a) panel (a.1) shows the applied bias across the 2D TI.
We apply a \Vd~pulse with a pulse width of 0.4 ns and an amplitude of 100 mV at $t=0.8$ ns.
Figure~\ref{f:damping} (a) panel (a.2) shows the normalized magnetization in the $\vec{z}$-direction ($S_{\rm {z}}/|M|$) as a function of time of the interface 2D FM for Gilbert damping parameter $\alpha=\,0.02,\,0.03,\,0.04,\,0.05$ for a pulse bias.
With the transition in applied \Vd, the normalized magnetization of the 2D magnet transitions from $1$ to $-1$.
We also observe that with increasing $\alpha$, the transition occurs faster, suggesting that the switching speed increases.

Figure~\ref{f:damping} (b) panels (b.1) and (b.2) show a pulse bias applied accross the 2D TI with a pulse width of 0.2 ns applied at 0.4 ns as a function of time, the normalized magnetization in the $\vec{z}$-direction ($S_{\rm{z}}/|M|$) as a function of time of the interface 2D FM for interface exchange parameter $J_{\mathrm{int}}=\,20,\,25,\,50,\,75,\,100$, and $150$ meV.
We observe that with increasing $J_{\mathrm{int}}$, the transition occurs faster up to $J_{\mathrm{int}}=50$ meV.
For $J_{\mathrm{int}}>50$ meV, we see that the magnetization does not switch smoothly, and the magnetization stabilizes for the entire duration of the pulse.
After the applied pulse returns to 0 mV, the magnetization still switches for all $J_{\mathrm{int}}>50$ meV, albeit with some delay.

Figure~\ref{f:damping} (c) shows the transition time for the applied pulse in Fig.~\ref{f:damping} (b) panel (b.1) as a function of $J_{\rm int}$.
We define the transition time as $t[S_{\mathrm{z}}<-0.7 M]-t[S_{\mathrm{z}}>0.7M]$.
We observe that the transition time goes down as a function of $J_{\rm int}$ up to $J_{\rm int}=50$ meV and then increases for $J_{\rm int}> 50$ meV.
Therefore, we find that an optimal value of interface exchange ($J_{\rm int}$) is required for obtaining a small transition time.

To further understand the reason for the low transition rate at higher $J_{\rm int}$, we compare the magnetization at $t=0.45$~ns in Fig.~\ref{f:damping} (d) for $J_{\rm int}=25$~meV and $J_{\rm int}=150$~meV.
The upper two panels show the magnetization of the 2D FM in the $\rm z$-direction ($S_{\rm{z}}/|M|$), and the lower two panels show the induced magnetization in the 2D TI (${m}_{\rm {z}}$).
We observe that for $J_{\rm int}=150$ meV, ${m}_{\rm z}$ is lower than the ${m}_{\rm z}$ for $J_{\rm int}=25$ meV.
Therefore, the lower magnetization of the TI results in a lower torque on the 2D FM, causing a delay in transition.
The reason for the lower magnetization due to higher $J_{\rm int}$ is because the magnet opens a bandgap in the 2D TI, resulting in a superposition of up and down spins~\cite{TI-FM-hetero1}.
The superposition of up and down spins in TI leads to a reduced magnetization (For further details, see Supplementary Fig. 1 and Fig. 2).
\begin{table}[ht]
\caption{$J$-parameters and anisotropies of experimental Cr-compounds} 
\centering
\resizebox{0.85\columnwidth}{!}{
\begin{tabular}{c c c c c c c} 
\toprule
Parameters  & $J_{\rm NN}$ &$J_{\rm NNN}$ & $J_{\rm NNNN}$& $\Delta_{\rm NN}$ &$\Delta_{\rm NNN}$ & $\Delta_{\rm NNNN}$ \\ [0.5ex]
                   & (meV) & (meV) & (meV) &  & & \\ [0.5ex]
\midrule
$\rm CrI_3$ & 2.21 & 0.75  & -  & 0.029  &  0.045 &-  \\
$\rm CrBr_3$ & 1.38 & 0.44  & -  & 0.010  &  0.012&-  \\
$\rm CrCl_3$ & 1.31 & 0.24  & -  & 0.001 &  0.008 &-  \\
$\rm CrGeTe_3$ &5.87 & -0.28  & 0.345 & 0.02 &  0.0 & 0.028  \\[1ex]
\bottomrule
\end{tabular}
}
\label{t:J_table}
\end{table}
\subsection{Impact of anisotropy and Cr-compounds}

We compare the transition time for the device in Fig.~\ref{f:switch_device} (a) as a function of nearest-neighbor anisotropy ($\Delta_{\rm NN}$) of the FM as shown in Fig.~\ref{f:Cr-compounds}. 
We perform the same sweep of $\Delta_{\rm NN}$ for 10 different starting configurations. The solid line shows the median and the shaded region shows the $25^{\rm th}$-$75^{\rm th}$ percentile.
We use $J_{\rm int}=0.025$ eV and $\alpha=0.05$ and nearest neighbor exchange for the 2D magnet to be $J_{\rm NN}=2.5$ meV.

We find that with increasing $\Delta_{\rm NN}$ the transition time increases. 
Therefore, for faster switching of the TI-FM device, it will be ideal to use materials with lower $\Delta_{\rm NN}$.
However, lower $\Delta_{\rm NN}$ results in a lower Curie temperature~\cite{Tiwari2021_Curie}.
Hence, the choice of the FM material should take into account the temperature of operation and the required switching time for the device.

For the device shown in Fig.~\ref{f:switch_device} (a), we use \CrI, \CrBr, \CrCl, and \CrGeTe~as the 2D magnetic material.
Although, many studies in previous works have used a parametric value of $\alpha$~\cite{Cr-magnetization} for Cr-compounds, recent reports on the measurement of intrinsic Gilbert damping of a sister compound \CrCl~have found a value of $\alpha=0.002$~\cite{alphaCrCl3}.
Unfortunately, there is no similar report for other Cr-compounds , therefore we have used an $\alpha$ of similar order, $\alpha=0.005$, for all the three Cr-compounds for a fair 
comparison.
The parameters used for modeling their magnetic structure of the Cr-compounds is shown in Table~\ref{t:J_table}.

Figure~\ref{f:Cr-compounds} (b) panel (b.1) shows the applied \Vd~and panel (b.2) shows the out-of-plane magnetization $S_{\rm z}$ for \CrI, \CrBr, \CrCl, and \CrGeTe.
We find that the magnetization of \CrI~is the most stable among the Cr-compounds followed by \CrGeTe, \CrBr, and \CrCl, respectively.
However, \CrI~is also the slowest to respond to a change in the applied bias and has the highest transition time.

Comparing the parameters for the Cr-compounds in Table~\ref{t:J_table}, we find that the transition time is inversely proportional to the nearest-neighbor anisotropy ($\Delta_{\rm NN}$).
The higher the anisotropy, the higher the transition time.
On the other hand, the stability of magnetization is directly proportional to the anisotropy.

\section{Conclusion}
We have presented a method to model spin-charge dynamics at a magnetic-topological insulator interface.
Our model is general and not limited to only the TI-FM interface.
Our method combines NEGF and TQMC to model the spin-charge dynamics.
The benefit of TQMC+NEGF over conventional methods such as LLG+QTBM or LLG+NEGF lies in the atomistic description of the spins, which allows one to implement the full exchange tensor for modelling the magnetic exchange interactions.

Using our method, we have theoretically investigated the spin-charge dynamics in a 2D TI-FM heterostructure where the size of the 2D FM was smaller than the 2D TI and placed on one of the edges of the 2D TI.
We have shown that by electrically biasing the 2D TI, it is possible to switch the magnetization of the 2D FM without destroying the edge state of the 2D TI.
The most important result of our work is that the switching of the 2D FM using 2D TI spin-torque can only be achieved efficiently in TI-FM material combinations that have an optimal interface exchange interaction $J_{\rm int}$.
We have also compared experimentally grown 2D Cr-compounds.
We have shown that the transition rate of magnetization significantly depends on the anisotropy, and low anisotropic Cr-compounds (\CrBr~and \CrCl) show faster switching in comparison to the higher anisotropic Cr-compounds (\CrI~and \CrGeTe).
Finally, there are recent reports of experimentally growing 2D FM-semiconductor heterostructures for spintronic devices~\cite{FM-hetero}, and given the device design presented in our paper is feasible to make experimentally, and we believe it should be explored experimentally. 

\section{Acknowledgements}
The project or effort depicted was or is sponsored by the Department of Defense, Defense Threat Reduction Agency.
The content of the information does not necessarily reflect the position or the policy of the federal government, and no official endorsement should be inferred.

This material is based upon work supported by the National Science Foundation under Grant No. 1802166.

We thank the Research Foundation Flanders (FWO) and the KU Leuven BOF Program (C14/18/074).

This work was supported by imec's Industrial Affiliation Program.

\section*{References}
\bibliography{bib}

\begin{thebibliography}{35}%
\makeatletter
\providecommand \@ifxundefined [1]{%
 \@ifx{#1\undefined}
}%
\providecommand \@ifnum [1]{%
 \ifnum #1\expandafter \@firstoftwo
 \else \expandafter \@secondoftwo
 \fi
}%
\providecommand \@ifx [1]{%
 \ifx #1\expandafter \@firstoftwo
 \else \expandafter \@secondoftwo
 \fi
}%
\providecommand \natexlab [1]{#1}%
\providecommand \enquote  [1]{``#1''}%
\providecommand \bibnamefont  [1]{#1}%
\providecommand \bibfnamefont [1]{#1}%
\providecommand \citenamefont [1]{#1}%
\providecommand \href@noop [0]{\@secondoftwo}%
\providecommand \href [0]{\begingroup \@sanitize@url \@href}%
\providecommand \@href[1]{\@@startlink{#1}\@@href}%
\providecommand \@@href[1]{\endgroup#1\@@endlink}%
\providecommand \@sanitize@url [0]{\catcode `\\12\catcode `\$12\catcode
  `\&12\catcode `\#12\catcode `\^12\catcode `\_12\catcode `\%12\relax}%
\providecommand \@@startlink[1]{}%
\providecommand \@@endlink[0]{}%
\providecommand \url  [0]{\begingroup\@sanitize@url \@url }%
\providecommand \@url [1]{\endgroup\@href {#1}{\urlprefix }}%
\providecommand \urlprefix  [0]{URL }%
\providecommand \Eprint [0]{\href }%
\providecommand \doibase [0]{https://doi.org/}%
\providecommand \selectlanguage [0]{\@gobble}%
\providecommand \bibinfo  [0]{\@secondoftwo}%
\providecommand \bibfield  [0]{\@secondoftwo}%
\providecommand \translation [1]{[#1]}%
\providecommand \BibitemOpen [0]{}%
\providecommand \bibitemStop [0]{}%
\providecommand \bibitemNoStop [0]{.\EOS\space}%
\providecommand \EOS [0]{\spacefactor3000\relax}%
\providecommand \BibitemShut  [1]{\csname bibitem#1\endcsname}%
\let\auto@bib@innerbib\@empty
\bibitem [{\citenamefont {Huang}\ \emph {et~al.}(2017)\citenamefont {Huang},
  \citenamefont {Clark}, \citenamefont {Navarro-Moratalla}, \citenamefont
  {Klein}, \citenamefont {Cheng}, \citenamefont {Seyler}, \citenamefont
  {Zhong}, \citenamefont {Schmidgall}, \citenamefont {McGuire}, \citenamefont
  {Cobden}, \citenamefont {Yao}, \citenamefont {Xiao}, \citenamefont
  {Jarillo-Herrero},\ and\ \citenamefont {Xu}}]{CrI_exp}%
  \BibitemOpen
  \bibfield  {author} {\bibinfo {author} {\bibfnamefont {B.}~\bibnamefont
  {Huang}}, \bibinfo {author} {\bibfnamefont {G.}~\bibnamefont {Clark}},
  \bibinfo {author} {\bibfnamefont {E.}~\bibnamefont {Navarro-Moratalla}},
  \bibinfo {author} {\bibfnamefont {D.~R.}\ \bibnamefont {Klein}}, \bibinfo
  {author} {\bibfnamefont {R.}~\bibnamefont {Cheng}}, \bibinfo {author}
  {\bibfnamefont {K.~L.}\ \bibnamefont {Seyler}}, \bibinfo {author}
  {\bibfnamefont {D.}~\bibnamefont {Zhong}}, \bibinfo {author} {\bibfnamefont
  {E.}~\bibnamefont {Schmidgall}}, \bibinfo {author} {\bibfnamefont {M.~A.}\
  \bibnamefont {McGuire}}, \bibinfo {author} {\bibfnamefont {D.~H.}\
  \bibnamefont {Cobden}}, \bibinfo {author} {\bibfnamefont {W.}~\bibnamefont
  {Yao}}, \bibinfo {author} {\bibfnamefont {D.}~\bibnamefont {Xiao}}, \bibinfo
  {author} {\bibfnamefont {P.}~\bibnamefont {Jarillo-Herrero}},\ and\ \bibinfo
  {author} {\bibfnamefont {X.}~\bibnamefont {Xu}},\ }\bibfield  {title}
  {\bibinfo {title} {Layer-dependent ferromagnetism in a van der waals crystal
  down to the monolayer limit},\ }\href@noop {} {\bibfield  {journal} {\bibinfo
   {journal} {Nature}\ }\textbf {\bibinfo {volume} {546}},\ \bibinfo {pages}
  {270 EP } (\bibinfo {year} {2017})}\BibitemShut {NoStop}%
\bibitem [{\citenamefont {Zhang}\ \emph {et~al.}(2019)\citenamefont {Zhang},
  \citenamefont {Shang}, \citenamefont {Jiang}, \citenamefont {Rasmita},
  \citenamefont {Gao},\ and\ \citenamefont {Yu}}]{CrBr_exp}%
  \BibitemOpen
  \bibfield  {author} {\bibinfo {author} {\bibfnamefont {Z.}~\bibnamefont
  {Zhang}}, \bibinfo {author} {\bibfnamefont {J.}~\bibnamefont {Shang}},
  \bibinfo {author} {\bibfnamefont {C.}~\bibnamefont {Jiang}}, \bibinfo
  {author} {\bibfnamefont {A.}~\bibnamefont {Rasmita}}, \bibinfo {author}
  {\bibfnamefont {W.}~\bibnamefont {Gao}},\ and\ \bibinfo {author}
  {\bibfnamefont {T.}~\bibnamefont {Yu}},\ }\bibfield  {title} {\bibinfo
  {title} {Direct photoluminescence probing of ferromagnetism in monolayer
  two-dimensional crbr3},\ }\href
  {https://doi.org/10.1021/acs.nanolett.9b00553} {\bibfield  {journal}
  {\bibinfo  {journal} {Nano Letters}\ }\textbf {\bibinfo {volume} {19}},\
  \bibinfo {pages} {3138} (\bibinfo {year} {2019})}\BibitemShut {NoStop}%
\bibitem [{\citenamefont {Gong}\ \emph {et~al.}(2017)\citenamefont {Gong},
  \citenamefont {Li}, \citenamefont {Li}, \citenamefont {Ji}, \citenamefont
  {Stern}, \citenamefont {Xia}, \citenamefont {Cao}, \citenamefont {Bao},
  \citenamefont {Wang}, \citenamefont {Wang}, \citenamefont {Qiu},
  \citenamefont {Cava}, \citenamefont {Louie}, \citenamefont {Xia},\ and\
  \citenamefont {Zhang}}]{CrGeTe_exp}%
  \BibitemOpen
  \bibfield  {author} {\bibinfo {author} {\bibfnamefont {C.}~\bibnamefont
  {Gong}}, \bibinfo {author} {\bibfnamefont {L.}~\bibnamefont {Li}}, \bibinfo
  {author} {\bibfnamefont {Z.}~\bibnamefont {Li}}, \bibinfo {author}
  {\bibfnamefont {H.}~\bibnamefont {Ji}}, \bibinfo {author} {\bibfnamefont
  {A.}~\bibnamefont {Stern}}, \bibinfo {author} {\bibfnamefont
  {Y.}~\bibnamefont {Xia}}, \bibinfo {author} {\bibfnamefont {T.}~\bibnamefont
  {Cao}}, \bibinfo {author} {\bibfnamefont {W.}~\bibnamefont {Bao}}, \bibinfo
  {author} {\bibfnamefont {C.}~\bibnamefont {Wang}}, \bibinfo {author}
  {\bibfnamefont {Y.}~\bibnamefont {Wang}}, \bibinfo {author} {\bibfnamefont
  {Z.~Q.}\ \bibnamefont {Qiu}}, \bibinfo {author} {\bibfnamefont {R.~J.}\
  \bibnamefont {Cava}}, \bibinfo {author} {\bibfnamefont {S.~G.}\ \bibnamefont
  {Louie}}, \bibinfo {author} {\bibfnamefont {J.}~\bibnamefont {Xia}},\ and\
  \bibinfo {author} {\bibfnamefont {X.}~\bibnamefont {Zhang}},\ }\bibfield
  {title} {\bibinfo {title} {Discovery of intrinsic ferromagnetism in
  two-dimensional van der waals crystals},\ }\href
  {https://doi.org/10.1038/nature22060} {\bibfield  {journal} {\bibinfo
  {journal} {Nature}\ }\textbf {\bibinfo {volume} {546}},\ \bibinfo {pages}
  {265 EP } (\bibinfo {year} {2017})}\BibitemShut {NoStop}%
\bibitem [{\citenamefont {Cortie}\ \emph {et~al.}(2020)\citenamefont {Cortie},
  \citenamefont {Causer}, \citenamefont {Rule}, \citenamefont {Fritzsche},
  \citenamefont {Kreuzpaintner},\ and\ \citenamefont {Klose}}]{sample21}%
  \BibitemOpen
  \bibfield  {author} {\bibinfo {author} {\bibfnamefont {D.~L.}\ \bibnamefont
  {Cortie}}, \bibinfo {author} {\bibfnamefont {G.~L.}\ \bibnamefont {Causer}},
  \bibinfo {author} {\bibfnamefont {K.~C.}\ \bibnamefont {Rule}}, \bibinfo
  {author} {\bibfnamefont {H.}~\bibnamefont {Fritzsche}}, \bibinfo {author}
  {\bibfnamefont {W.}~\bibnamefont {Kreuzpaintner}},\ and\ \bibinfo {author}
  {\bibfnamefont {F.}~\bibnamefont {Klose}},\ }\bibfield  {title} {\bibinfo
  {title} {Two-dimensional magnets: Forgotten history and recent progress
  towards spintronic applications},\ }\href
  {https://doi.org/https://doi.org/10.1002/adfm.201901414} {\bibfield
  {journal} {\bibinfo  {journal} {Advanced Functional Materials}\ }\textbf
  {\bibinfo {volume} {30}},\ \bibinfo {pages} {1901414} (\bibinfo {year}
  {2020})}\BibitemShut {NoStop}%
\bibitem [{\citenamefont {Wang}\ \emph {et~al.}(2018)\citenamefont {Wang},
  \citenamefont {Guti{\'e}rrez-Lezama}, \citenamefont {Ubrig}, \citenamefont
  {Kroner}, \citenamefont {Gibertini}, \citenamefont {Taniguchi}, \citenamefont
  {Watanabe}, \citenamefont {Imamo{\u{g}}lu}, \citenamefont {Giannini},\ and\
  \citenamefont {Morpurgo}}]{Nat_comm}%
  \BibitemOpen
  \bibfield  {author} {\bibinfo {author} {\bibfnamefont {Z.}~\bibnamefont
  {Wang}}, \bibinfo {author} {\bibfnamefont {I.}~\bibnamefont
  {Guti{\'e}rrez-Lezama}}, \bibinfo {author} {\bibfnamefont {N.}~\bibnamefont
  {Ubrig}}, \bibinfo {author} {\bibfnamefont {M.}~\bibnamefont {Kroner}},
  \bibinfo {author} {\bibfnamefont {M.}~\bibnamefont {Gibertini}}, \bibinfo
  {author} {\bibfnamefont {T.}~\bibnamefont {Taniguchi}}, \bibinfo {author}
  {\bibfnamefont {K.}~\bibnamefont {Watanabe}}, \bibinfo {author}
  {\bibfnamefont {A.}~\bibnamefont {Imamo{\u{g}}lu}}, \bibinfo {author}
  {\bibfnamefont {E.}~\bibnamefont {Giannini}},\ and\ \bibinfo {author}
  {\bibfnamefont {A.~F.}\ \bibnamefont {Morpurgo}},\ }\bibfield  {title}
  {\bibinfo {title} {Very large tunneling magnetoresistance in layered magnetic
  semiconductor cri3},\ }\href {https://doi.org/10.1038/s41467-018-04953-8}
  {\bibfield  {journal} {\bibinfo  {journal} {Nature Communications}\ }\textbf
  {\bibinfo {volume} {9}},\ \bibinfo {pages} {2516} (\bibinfo {year}
  {2018})}\BibitemShut {NoStop}%
\bibitem [{\citenamefont {Zhong}\ \emph
  {et~al.}(2017{\natexlab{a}})\citenamefont {Zhong}, \citenamefont {Seyler},
  \citenamefont {Linpeng}, \citenamefont {Cheng}, \citenamefont {Sivadas},
  \citenamefont {Huang}, \citenamefont {Schmidgall}, \citenamefont {Taniguchi},
  \citenamefont {Watanabe}, \citenamefont {McGuire}, \citenamefont {Yao},
  \citenamefont {Xiao}, \citenamefont {Fu},\ and\ \citenamefont
  {Xu}}]{sample37}%
  \BibitemOpen
  \bibfield  {author} {\bibinfo {author} {\bibfnamefont {D.}~\bibnamefont
  {Zhong}}, \bibinfo {author} {\bibfnamefont {K.~L.}\ \bibnamefont {Seyler}},
  \bibinfo {author} {\bibfnamefont {X.}~\bibnamefont {Linpeng}}, \bibinfo
  {author} {\bibfnamefont {R.}~\bibnamefont {Cheng}}, \bibinfo {author}
  {\bibfnamefont {N.}~\bibnamefont {Sivadas}}, \bibinfo {author} {\bibfnamefont
  {B.}~\bibnamefont {Huang}}, \bibinfo {author} {\bibfnamefont
  {E.}~\bibnamefont {Schmidgall}}, \bibinfo {author} {\bibfnamefont
  {T.}~\bibnamefont {Taniguchi}}, \bibinfo {author} {\bibfnamefont
  {K.}~\bibnamefont {Watanabe}}, \bibinfo {author} {\bibfnamefont {M.~A.}\
  \bibnamefont {McGuire}}, \bibinfo {author} {\bibfnamefont {W.}~\bibnamefont
  {Yao}}, \bibinfo {author} {\bibfnamefont {D.}~\bibnamefont {Xiao}}, \bibinfo
  {author} {\bibfnamefont {K.-M.~C.}\ \bibnamefont {Fu}},\ and\ \bibinfo
  {author} {\bibfnamefont {X.}~\bibnamefont {Xu}},\ }\bibfield  {title}
  {\bibinfo {title} {Van der waals engineering of ferromagnetic semiconductor
  heterostructures for spin and valleytronics},\ }\bibfield  {journal}
  {\bibinfo  {journal} {Science Advances}\ }\textbf {\bibinfo {volume} {3}},\
  \href {https://doi.org/10.1126/sciadv.1603113} {10.1126/sciadv.1603113}
  (\bibinfo {year} {2017}{\natexlab{a}})\BibitemShut {NoStop}%
\bibitem [{\citenamefont {Amoroso}\ \emph {et~al.}(2020)\citenamefont
  {Amoroso}, \citenamefont {Barone},\ and\ \citenamefont
  {Picozzi}}]{Nat_comm2}%
  \BibitemOpen
  \bibfield  {author} {\bibinfo {author} {\bibfnamefont {D.}~\bibnamefont
  {Amoroso}}, \bibinfo {author} {\bibfnamefont {P.}~\bibnamefont {Barone}},\
  and\ \bibinfo {author} {\bibfnamefont {S.}~\bibnamefont {Picozzi}},\
  }\bibfield  {title} {\bibinfo {title} {Spontaneous skyrmionic lattice from
  anisotropic symmetric exchange in a ni-halide monolayer},\ }\href
  {https://doi.org/10.1038/s41467-020-19535-w} {\bibfield  {journal} {\bibinfo
  {journal} {Nature Communications}\ }\textbf {\bibinfo {volume} {11}},\
  \bibinfo {pages} {5784} (\bibinfo {year} {2020})}\BibitemShut {NoStop}%
\bibitem [{\citenamefont {Tong}\ \emph {et~al.}(2018)\citenamefont {Tong},
  \citenamefont {Liu}, \citenamefont {Xiao},\ and\ \citenamefont
  {Yao}}]{sample22}%
  \BibitemOpen
  \bibfield  {author} {\bibinfo {author} {\bibfnamefont {Q.}~\bibnamefont
  {Tong}}, \bibinfo {author} {\bibfnamefont {F.}~\bibnamefont {Liu}}, \bibinfo
  {author} {\bibfnamefont {J.}~\bibnamefont {Xiao}},\ and\ \bibinfo {author}
  {\bibfnamefont {W.}~\bibnamefont {Yao}},\ }\bibfield  {title} {\bibinfo
  {title} {Skyrmions in the moir{\'e} of van der waals 2d magnets},\ }\href
  {https://doi.org/10.1021/acs.nanolett.8b03315} {\bibfield  {journal}
  {\bibinfo  {journal} {Nano Letters}\ }\textbf {\bibinfo {volume} {18}},\
  \bibinfo {pages} {7194} (\bibinfo {year} {2018})}\BibitemShut {NoStop}%
\bibitem [{\citenamefont {Fan}\ \emph {et~al.}(2014)\citenamefont {Fan},
  \citenamefont {Upadhyaya}, \citenamefont {Kou}, \citenamefont {Lang},
  \citenamefont {Takei}, \citenamefont {Wang}, \citenamefont {Tang},
  \citenamefont {He}, \citenamefont {Chang}, \citenamefont {Montazeri},
  \citenamefont {Yu}, \citenamefont {Jiang}, \citenamefont {Nie}, \citenamefont
  {Schwartz}, \citenamefont {Tserkovnyak},\ and\ \citenamefont
  {Wang}}]{TI-FM1}%
  \BibitemOpen
  \bibfield  {author} {\bibinfo {author} {\bibfnamefont {Y.}~\bibnamefont
  {Fan}}, \bibinfo {author} {\bibfnamefont {P.}~\bibnamefont {Upadhyaya}},
  \bibinfo {author} {\bibfnamefont {X.}~\bibnamefont {Kou}}, \bibinfo {author}
  {\bibfnamefont {M.}~\bibnamefont {Lang}}, \bibinfo {author} {\bibfnamefont
  {S.}~\bibnamefont {Takei}}, \bibinfo {author} {\bibfnamefont
  {Z.}~\bibnamefont {Wang}}, \bibinfo {author} {\bibfnamefont {J.}~\bibnamefont
  {Tang}}, \bibinfo {author} {\bibfnamefont {L.}~\bibnamefont {He}}, \bibinfo
  {author} {\bibfnamefont {L.-T.}\ \bibnamefont {Chang}}, \bibinfo {author}
  {\bibfnamefont {M.}~\bibnamefont {Montazeri}}, \bibinfo {author}
  {\bibfnamefont {G.}~\bibnamefont {Yu}}, \bibinfo {author} {\bibfnamefont
  {W.}~\bibnamefont {Jiang}}, \bibinfo {author} {\bibfnamefont
  {T.}~\bibnamefont {Nie}}, \bibinfo {author} {\bibfnamefont {R.~N.}\
  \bibnamefont {Schwartz}}, \bibinfo {author} {\bibfnamefont {Y.}~\bibnamefont
  {Tserkovnyak}},\ and\ \bibinfo {author} {\bibfnamefont {K.~L.}\ \bibnamefont
  {Wang}},\ }\bibfield  {title} {\bibinfo {title} {Magnetization switching
  through giant spin--orbit torque in a magnetically doped topological
  insulator heterostructure},\ }\href {https://doi.org/10.1038/nmat3973}
  {\bibfield  {journal} {\bibinfo  {journal} {Nature Materials}\ }\textbf
  {\bibinfo {volume} {13}},\ \bibinfo {pages} {699} (\bibinfo {year}
  {2014})}\BibitemShut {NoStop}%
\bibitem [{\citenamefont {Wang}\ \emph {et~al.}(2017)\citenamefont {Wang},
  \citenamefont {Zhu}, \citenamefont {Wu}, \citenamefont {Yang}, \citenamefont
  {Yu}, \citenamefont {Ramaswamy}, \citenamefont {Mishra}, \citenamefont {Shi},
  \citenamefont {Elyasi}, \citenamefont {Teo}, \citenamefont {Wu},\ and\
  \citenamefont {Yang}}]{TI-FM2}%
  \BibitemOpen
  \bibfield  {author} {\bibinfo {author} {\bibfnamefont {Y.}~\bibnamefont
  {Wang}}, \bibinfo {author} {\bibfnamefont {D.}~\bibnamefont {Zhu}}, \bibinfo
  {author} {\bibfnamefont {Y.}~\bibnamefont {Wu}}, \bibinfo {author}
  {\bibfnamefont {Y.}~\bibnamefont {Yang}}, \bibinfo {author} {\bibfnamefont
  {J.}~\bibnamefont {Yu}}, \bibinfo {author} {\bibfnamefont {R.}~\bibnamefont
  {Ramaswamy}}, \bibinfo {author} {\bibfnamefont {R.}~\bibnamefont {Mishra}},
  \bibinfo {author} {\bibfnamefont {S.}~\bibnamefont {Shi}}, \bibinfo {author}
  {\bibfnamefont {M.}~\bibnamefont {Elyasi}}, \bibinfo {author} {\bibfnamefont
  {K.-L.}\ \bibnamefont {Teo}}, \bibinfo {author} {\bibfnamefont
  {Y.}~\bibnamefont {Wu}},\ and\ \bibinfo {author} {\bibfnamefont
  {H.}~\bibnamefont {Yang}},\ }\bibfield  {title} {\bibinfo {title} {Room
  temperature magnetization switching in topological insulator-ferromagnet
  heterostructures by spin-orbit torques},\ }\href
  {https://doi.org/10.1038/s41467-017-01583-4} {\bibfield  {journal} {\bibinfo
  {journal} {Nature Communications}\ }\textbf {\bibinfo {volume} {8}},\
  \bibinfo {pages} {1364} (\bibinfo {year} {2017})}\BibitemShut {NoStop}%
\bibitem [{\citenamefont {Han}\ \emph {et~al.}(2017)\citenamefont {Han},
  \citenamefont {Richardella}, \citenamefont {Siddiqui}, \citenamefont
  {Finley}, \citenamefont {Samarth},\ and\ \citenamefont {Liu}}]{TI-FM3}%
  \BibitemOpen
  \bibfield  {author} {\bibinfo {author} {\bibfnamefont {J.}~\bibnamefont
  {Han}}, \bibinfo {author} {\bibfnamefont {A.}~\bibnamefont {Richardella}},
  \bibinfo {author} {\bibfnamefont {S.~A.}\ \bibnamefont {Siddiqui}}, \bibinfo
  {author} {\bibfnamefont {J.}~\bibnamefont {Finley}}, \bibinfo {author}
  {\bibfnamefont {N.}~\bibnamefont {Samarth}},\ and\ \bibinfo {author}
  {\bibfnamefont {L.}~\bibnamefont {Liu}},\ }\bibfield  {title} {\bibinfo
  {title} {Room-temperature spin-orbit torque switching induced by a
  topological insulator},\ }\href
  {https://doi.org/10.1103/PhysRevLett.119.077702} {\bibfield  {journal}
  {\bibinfo  {journal} {Phys. Rev. Lett.}\ }\textbf {\bibinfo {volume} {119}},\
  \bibinfo {pages} {077702} (\bibinfo {year} {2017})}\BibitemShut {NoStop}%
\bibitem [{\citenamefont {Fan}\ \emph {et~al.}(2016)\citenamefont {Fan},
  \citenamefont {Kou}, \citenamefont {Upadhyaya}, \citenamefont {Shao},
  \citenamefont {Pan}, \citenamefont {Lang}, \citenamefont {Che}, \citenamefont
  {Tang}, \citenamefont {Montazeri}, \citenamefont {Murata}, \citenamefont
  {Chang}, \citenamefont {Akyol}, \citenamefont {Yu}, \citenamefont {Nie},
  \citenamefont {Wong}, \citenamefont {Liu}, \citenamefont {Wang},
  \citenamefont {Tserkovnyak},\ and\ \citenamefont {Wang}}]{TI-FM4}%
  \BibitemOpen
  \bibfield  {author} {\bibinfo {author} {\bibfnamefont {Y.}~\bibnamefont
  {Fan}}, \bibinfo {author} {\bibfnamefont {X.}~\bibnamefont {Kou}}, \bibinfo
  {author} {\bibfnamefont {P.}~\bibnamefont {Upadhyaya}}, \bibinfo {author}
  {\bibfnamefont {Q.}~\bibnamefont {Shao}}, \bibinfo {author} {\bibfnamefont
  {L.}~\bibnamefont {Pan}}, \bibinfo {author} {\bibfnamefont {M.}~\bibnamefont
  {Lang}}, \bibinfo {author} {\bibfnamefont {X.}~\bibnamefont {Che}}, \bibinfo
  {author} {\bibfnamefont {J.}~\bibnamefont {Tang}}, \bibinfo {author}
  {\bibfnamefont {M.}~\bibnamefont {Montazeri}}, \bibinfo {author}
  {\bibfnamefont {K.}~\bibnamefont {Murata}}, \bibinfo {author} {\bibfnamefont
  {L.-T.}\ \bibnamefont {Chang}}, \bibinfo {author} {\bibfnamefont
  {M.}~\bibnamefont {Akyol}}, \bibinfo {author} {\bibfnamefont
  {G.}~\bibnamefont {Yu}}, \bibinfo {author} {\bibfnamefont {T.}~\bibnamefont
  {Nie}}, \bibinfo {author} {\bibfnamefont {K.~L.}\ \bibnamefont {Wong}},
  \bibinfo {author} {\bibfnamefont {J.}~\bibnamefont {Liu}}, \bibinfo {author}
  {\bibfnamefont {Y.}~\bibnamefont {Wang}}, \bibinfo {author} {\bibfnamefont
  {Y.}~\bibnamefont {Tserkovnyak}},\ and\ \bibinfo {author} {\bibfnamefont
  {K.~L.}\ \bibnamefont {Wang}},\ }\bibfield  {title} {\bibinfo {title}
  {Electric-field control of spin--orbit torque in a magnetically doped
  topological insulator},\ }\href {https://doi.org/10.1038/nnano.2015.294}
  {\bibfield  {journal} {\bibinfo  {journal} {Nature Nanotechnology}\ }\textbf
  {\bibinfo {volume} {11}},\ \bibinfo {pages} {352} (\bibinfo {year}
  {2016})}\BibitemShut {NoStop}%
\bibitem [{\citenamefont {Rachel}\ and\ \citenamefont {Ezawa}(2014)}]{TI-FM5}%
  \BibitemOpen
  \bibfield  {author} {\bibinfo {author} {\bibfnamefont {S.}~\bibnamefont
  {Rachel}}\ and\ \bibinfo {author} {\bibfnamefont {M.}~\bibnamefont {Ezawa}},\
  }\bibfield  {title} {\bibinfo {title} {Giant magnetoresistance and perfect
  spin filter in silicene, germanene, and stanene},\ }\href
  {https://doi.org/10.1103/PhysRevB.89.195303} {\bibfield  {journal} {\bibinfo
  {journal} {Phys. Rev. B}\ }\textbf {\bibinfo {volume} {89}},\ \bibinfo
  {pages} {195303} (\bibinfo {year} {2014})}\BibitemShut {NoStop}%
\bibitem [{\citenamefont {Van~Dyke}\ and\ \citenamefont
  {Morr}(2017)}]{TI-FM-hetero1}%
  \BibitemOpen
  \bibfield  {author} {\bibinfo {author} {\bibfnamefont {J.~S.}\ \bibnamefont
  {Van~Dyke}}\ and\ \bibinfo {author} {\bibfnamefont {D.~K.}\ \bibnamefont
  {Morr}},\ }\bibfield  {title} {\bibinfo {title} {Effects of defects and
  dephasing on charge and spin currents in two-dimensional topological
  insulators},\ }\href {https://doi.org/10.1103/PhysRevB.95.045151} {\bibfield
  {journal} {\bibinfo  {journal} {Phys. Rev. B}\ }\textbf {\bibinfo {volume}
  {95}},\ \bibinfo {pages} {045151} (\bibinfo {year} {2017})}\BibitemShut
  {NoStop}%
\bibitem [{\citenamefont {Hou}\ \emph {et~al.}(2019)\citenamefont {Hou},
  \citenamefont {Kim},\ and\ \citenamefont {Wu}}]{TI-FM-hetero2}%
  \BibitemOpen
  \bibfield  {author} {\bibinfo {author} {\bibfnamefont {Y.}~\bibnamefont
  {Hou}}, \bibinfo {author} {\bibfnamefont {J.}~\bibnamefont {Kim}},\ and\
  \bibinfo {author} {\bibfnamefont {R.}~\bibnamefont {Wu}},\ }\bibfield
  {title} {\bibinfo {title} {Magnetizing topological surface states of {$\rm
  Bi_2Se_3$} with a {$\rm CrI_3$} monolayer},\ }\href
  {https://doi.org/10.1126/sciadv.aaw1874} {\bibfield  {journal} {\bibinfo
  {journal} {Science Advances}\ }\textbf {\bibinfo {volume} {5}},\ \bibinfo
  {pages} {eaaw1874} (\bibinfo {year} {2019})}\BibitemShut {NoStop}%
\bibitem [{\citenamefont {Osca}\ and\ \citenamefont
  {Sor\'ee}(2020)}]{Javier2020}%
  \BibitemOpen
  \bibfield  {author} {\bibinfo {author} {\bibfnamefont {J.}~\bibnamefont
  {Osca}}\ and\ \bibinfo {author} {\bibfnamefont {B.}~\bibnamefont {Sor\'ee}},\
  }\bibfield  {title} {\bibinfo {title} {Skyrmion spin transfer torque due to
  current confined in a nanowire},\ }\href
  {https://doi.org/10.1103/PhysRevB.102.125436} {\bibfield  {journal} {\bibinfo
   {journal} {Phys. Rev. B}\ }\textbf {\bibinfo {volume} {102}},\ \bibinfo
  {pages} {125436} (\bibinfo {year} {2020})}\BibitemShut {NoStop}%
\bibitem [{\citenamefont {Osca}\ and\ \citenamefont
  {Sorée}(2021)}]{Javier-JAP}%
  \BibitemOpen
  \bibfield  {author} {\bibinfo {author} {\bibfnamefont {J.}~\bibnamefont
  {Osca}}\ and\ \bibinfo {author} {\bibfnamefont {B.}~\bibnamefont {Sorée}},\
  }\bibfield  {title} {\bibinfo {title} {Torque field and skyrmion motion by
  spin transfer torque in a quasi-2d interface in presence of strong
  spin–orbit interaction},\ }\href {https://doi.org/10.1063/5.0063887}
  {\bibfield  {journal} {\bibinfo  {journal} {Journal of Applied Physics}\
  }\textbf {\bibinfo {volume} {130}},\ \bibinfo {pages} {133903} (\bibinfo
  {year} {2021})}\BibitemShut {NoStop}%
\bibitem [{\citenamefont {Osca}\ \emph {et~al.}(2021)\citenamefont {Osca},
  \citenamefont {Moors}, \citenamefont {Sor{\'{e}}e},\ and\ \citenamefont
  {Serra}}]{Javier-3D-TI}%
  \BibitemOpen
  \bibfield  {author} {\bibinfo {author} {\bibfnamefont {J.}~\bibnamefont
  {Osca}}, \bibinfo {author} {\bibfnamefont {K.}~\bibnamefont {Moors}},
  \bibinfo {author} {\bibfnamefont {B.}~\bibnamefont {Sor{\'{e}}e}},\ and\
  \bibinfo {author} {\bibfnamefont {L.}~\bibnamefont {Serra}},\ }\bibfield
  {title} {\bibinfo {title} {Fabryp{\'{e}}rot interferometry with gate-tunable
  3d topological insulator nanowires},\ }\href
  {https://doi.org/10.1088/1361-6528/ac1633} {\ \textbf {\bibinfo {volume}
  {32}},\ \bibinfo {pages} {435002} (\bibinfo {year} {2021})}\BibitemShut
  {NoStop}%
\bibitem [{\citenamefont {Landau}\ and\ \citenamefont {Lifshitz}(1992)}]{LLG}%
  \BibitemOpen
  \bibfield  {author} {\bibinfo {author} {\bibfnamefont {L.}~\bibnamefont
  {Landau}}\ and\ \bibinfo {author} {\bibfnamefont {E.}~\bibnamefont
  {Lifshitz}},\ }\bibfield  {title} {\bibinfo {title} {On the theory of the
  dispersion of magnetic permeability in ferromagnetic bodies reprinted from
  physikalische zeitschrift der sowjetunion 8, part 2, 153, 1935.},\ }in\ \href
  {https://doi.org/https://doi.org/10.1016/B978-0-08-036364-6.50008-9} {\emph
  {\bibinfo {booktitle} {Perspectives in Theoretical Physics}}},\ \bibinfo
  {editor} {edited by\ \bibinfo {editor} {\bibfnamefont {L.}~\bibnamefont
  {Pitaevski}}}\ (\bibinfo  {publisher} {Pergamon},\ \bibinfo {address}
  {Amsterdam},\ \bibinfo {year} {1992})\ pp.\ \bibinfo {pages}
  {51--65}\BibitemShut {NoStop}%
\bibitem [{\citenamefont {Petrovi\ifmmode~\acute{c}\else \'{c}\fi{}}\ \emph
  {et~al.}(2018)\citenamefont {Petrovi\ifmmode~\acute{c}\else \'{c}\fi{}},
  \citenamefont {Popescu}, \citenamefont {Bajpai}, \citenamefont
  {Plech\'a\ifmmode~\check{c}\else \v{c}\fi{}},\ and\ \citenamefont
  {Nikoli\ifmmode~\acute{c}\else \'{c}\fi{}}}]{NEGF-LLG}%
  \BibitemOpen
  \bibfield  {author} {\bibinfo {author} {\bibfnamefont {M.~D.}\ \bibnamefont
  {Petrovi\ifmmode~\acute{c}\else \'{c}\fi{}}}, \bibinfo {author}
  {\bibfnamefont {B.~S.}\ \bibnamefont {Popescu}}, \bibinfo {author}
  {\bibfnamefont {U.}~\bibnamefont {Bajpai}}, \bibinfo {author} {\bibfnamefont
  {P.}~\bibnamefont {Plech\'a\ifmmode~\check{c}\else \v{c}\fi{}}},\ and\
  \bibinfo {author} {\bibfnamefont {B.~K.}\ \bibnamefont
  {Nikoli\ifmmode~\acute{c}\else \'{c}\fi{}}},\ }\bibfield  {title} {\bibinfo
  {title} {Spin and charge pumping by a steady or pulse-current-driven magnetic
  domain wall: A self-consistent multiscale time-dependent quantum-classical
  hybrid approach},\ }\href {https://doi.org/10.1103/PhysRevApplied.10.054038}
  {\bibfield  {journal} {\bibinfo  {journal} {Phys. Rev. Applied}\ }\textbf
  {\bibinfo {volume} {10}},\ \bibinfo {pages} {054038} (\bibinfo {year}
  {2018})}\BibitemShut {NoStop}%
\bibitem [{\citenamefont {Tiwari}\ \emph
  {et~al.}(2021{\natexlab{a}})\citenamefont {Tiwari}, \citenamefont {Vanherck},
  \citenamefont {Van~de Put}, \citenamefont {Vandenberghe},\ and\ \citenamefont
  {Sor\'ee}}]{Tiwari2021_Curie}%
  \BibitemOpen
  \bibfield  {author} {\bibinfo {author} {\bibfnamefont {S.}~\bibnamefont
  {Tiwari}}, \bibinfo {author} {\bibfnamefont {J.}~\bibnamefont {Vanherck}},
  \bibinfo {author} {\bibfnamefont {M.~L.}\ \bibnamefont {Van~de Put}},
  \bibinfo {author} {\bibfnamefont {W.~G.}\ \bibnamefont {Vandenberghe}},\ and\
  \bibinfo {author} {\bibfnamefont {B.}~\bibnamefont {Sor\'ee}},\ }\bibfield
  {title} {\bibinfo {title} {Computing curie temperature of two-dimensional
  ferromagnets in the presence of exchange anisotropy},\ }\href
  {https://doi.org/10.1103/PhysRevResearch.3.043024} {\bibfield  {journal}
  {\bibinfo  {journal} {Phys. Rev. Research}\ }\textbf {\bibinfo {volume}
  {3}},\ \bibinfo {pages} {043024} (\bibinfo {year}
  {2021}{\natexlab{a}})}\BibitemShut {NoStop}%
\bibitem [{\citenamefont {Tiwari}\ \emph
  {et~al.}(2021{\natexlab{b}})\citenamefont {Tiwari}, \citenamefont {Van~de
  Put}, \citenamefont {Sor\'ee},\ and\ \citenamefont
  {Vandenberghe}}]{my-paper}%
  \BibitemOpen
  \bibfield  {author} {\bibinfo {author} {\bibfnamefont {S.}~\bibnamefont
  {Tiwari}}, \bibinfo {author} {\bibfnamefont {M.~L.}\ \bibnamefont {Van~de
  Put}}, \bibinfo {author} {\bibfnamefont {B.}~\bibnamefont {Sor\'ee}},\ and\
  \bibinfo {author} {\bibfnamefont {W.~G.}\ \bibnamefont {Vandenberghe}},\
  }\bibfield  {title} {\bibinfo {title} {Critical behavior of the ferromagnets
  {${\mathrm{CrI}}_{3}$}, {${\mathrm{CrBr}}_{3}$}, and
  {${\mathrm{CrGeTe}}_{3}$} and the antiferromagnet {${\mathrm{FeCl}}_{2}$}: A
  detailed first-principles study},\ }\href
  {https://doi.org/10.1103/PhysRevB.103.014432} {\bibfield  {journal} {\bibinfo
   {journal} {Phys. Rev. B}\ }\textbf {\bibinfo {volume} {103}},\ \bibinfo
  {pages} {014432} (\bibinfo {year} {2021}{\natexlab{b}})}\BibitemShut
  {NoStop}%
\bibitem [{\citenamefont {Vanherck}\ \emph {et~al.}(2018)\citenamefont
  {Vanherck}, \citenamefont {Sor{\'{e}}e},\ and\ \citenamefont
  {Magnus}}]{Vanherck18}%
  \BibitemOpen
  \bibfield  {author} {\bibinfo {author} {\bibfnamefont {J.}~\bibnamefont
  {Vanherck}}, \bibinfo {author} {\bibfnamefont {B.}~\bibnamefont
  {Sor{\'{e}}e}},\ and\ \bibinfo {author} {\bibfnamefont {W.}~\bibnamefont
  {Magnus}},\ }\bibfield  {title} {\bibinfo {title} {Anisotropic bulk and
  planar heisenberg ferromagnets in uniform, arbitrarily oriented magnetic
  fields},\ }\href {https://doi.org/10.1088/1361-648x/aac65f} {\bibfield
  {journal} {\bibinfo  {journal} {J. Phys.: Condens. Matter}\ }\textbf
  {\bibinfo {volume} {30}},\ \bibinfo {pages} {275801} (\bibinfo {year}
  {2018})}\BibitemShut {NoStop}%
\bibitem [{\citenamefont {Vanherck}\ \emph {et~al.}(2020)\citenamefont
  {Vanherck}, \citenamefont {Bacaksiz}, \citenamefont {Sor{\'{e}}e},
  \citenamefont {Milo{\v{s}}evi{\'{c}}},\ and\ \citenamefont
  {Magnus}}]{Vanherck20}%
  \BibitemOpen
  \bibfield  {author} {\bibinfo {author} {\bibfnamefont {J.}~\bibnamefont
  {Vanherck}}, \bibinfo {author} {\bibfnamefont {C.}~\bibnamefont {Bacaksiz}},
  \bibinfo {author} {\bibfnamefont {B.}~\bibnamefont {Sor{\'{e}}e}}, \bibinfo
  {author} {\bibfnamefont {M.~V.}\ \bibnamefont {Milo{\v{s}}evi{\'{c}}}},\ and\
  \bibinfo {author} {\bibfnamefont {W.}~\bibnamefont {Magnus}},\ }\bibfield
  {title} {\bibinfo {title} {{2D ferromagnetism at finite temperatures under
  quantum scrutiny}},\ }\href {https://doi.org/10.1063/5.0015619} {\bibfield
  {journal} {\bibinfo  {journal} {Appl. Phys. Lett.}\ }\textbf {\bibinfo
  {volume} {117}},\ \bibinfo {pages} {052401} (\bibinfo {year}
  {2020})}\BibitemShut {NoStop}%
\bibitem [{\citenamefont {Tiwari}\ \emph
  {et~al.}(2021{\natexlab{c}})\citenamefont {Tiwari}, \citenamefont {Van~de
  Put}, \citenamefont {Sor{\'e}e},\ and\ \citenamefont
  {Vandenberghe}}]{Tiwari2021}%
  \BibitemOpen
  \bibfield  {author} {\bibinfo {author} {\bibfnamefont {S.}~\bibnamefont
  {Tiwari}}, \bibinfo {author} {\bibfnamefont {M.~L.}\ \bibnamefont {Van~de
  Put}}, \bibinfo {author} {\bibfnamefont {B.}~\bibnamefont {Sor{\'e}e}},\ and\
  \bibinfo {author} {\bibfnamefont {W.~G.}\ \bibnamefont {Vandenberghe}},\
  }\bibfield  {title} {\bibinfo {title} {Magnetic order and critical
  temperature of substitutionally doped transition metal dichalcogenide
  monolayers},\ }\href {https://doi.org/10.1038/s41699-021-00233-0} {\bibfield
  {journal} {\bibinfo  {journal} {npj 2D Materials and Applications}\ }\textbf
  {\bibinfo {volume} {5}},\ \bibinfo {pages} {54} (\bibinfo {year}
  {2021}{\natexlab{c}})}\BibitemShut {NoStop}%
\bibitem [{\citenamefont {Zhang}\ \emph {et~al.}(2020)\citenamefont {Zhang},
  \citenamefont {Li}, \citenamefont {Weber}, \citenamefont {Goldberger},
  \citenamefont {Mak},\ and\ \citenamefont {Shan}}]{LLG-bad}%
  \BibitemOpen
  \bibfield  {author} {\bibinfo {author} {\bibfnamefont {X.-X.}\ \bibnamefont
  {Zhang}}, \bibinfo {author} {\bibfnamefont {L.}~\bibnamefont {Li}}, \bibinfo
  {author} {\bibfnamefont {D.}~\bibnamefont {Weber}}, \bibinfo {author}
  {\bibfnamefont {J.}~\bibnamefont {Goldberger}}, \bibinfo {author}
  {\bibfnamefont {K.~F.}\ \bibnamefont {Mak}},\ and\ \bibinfo {author}
  {\bibfnamefont {J.}~\bibnamefont {Shan}},\ }\bibfield  {title} {\bibinfo
  {title} {Gate-tunable spin waves in antiferromagnetic atomic bilayers},\
  }\href {https://doi.org/10.1038/s41563-020-0713-9} {\bibfield  {journal}
  {\bibinfo  {journal} {Nature Materials}\ }\textbf {\bibinfo {volume} {19}},\
  \bibinfo {pages} {838} (\bibinfo {year} {2020})}\BibitemShut {NoStop}%
\bibitem [{\citenamefont {Lado}\ and\ \citenamefont
  {Fern{\'{a}}ndez-Rossier}(2017)}]{Lado_2017}%
  \BibitemOpen
  \bibfield  {author} {\bibinfo {author} {\bibfnamefont {J.~L.}\ \bibnamefont
  {Lado}}\ and\ \bibinfo {author} {\bibfnamefont {J.}~\bibnamefont
  {Fern{\'{a}}ndez-Rossier}},\ }\bibfield  {title} {\bibinfo {title} {On the
  origin of magnetic anisotropy in two dimensional {CrI}3},\ }\href
  {https://doi.org/10.1088/2053-1583/aa75ed} {\bibfield  {journal} {\bibinfo
  {journal} {2D Materials}\ }\textbf {\bibinfo {volume} {4}},\ \bibinfo {pages}
  {035002} (\bibinfo {year} {2017})}\BibitemShut {NoStop}%
\bibitem [{\citenamefont {Chubykalo}\ \emph {et~al.}(2003)\citenamefont
  {Chubykalo}, \citenamefont {Nowak}, \citenamefont {Smirnov-Rueda},
  \citenamefont {Wongsam}, \citenamefont {Chantrell},\ and\ \citenamefont
  {Gonzalez}}]{TQMC1}%
  \BibitemOpen
  \bibfield  {author} {\bibinfo {author} {\bibfnamefont {O.}~\bibnamefont
  {Chubykalo}}, \bibinfo {author} {\bibfnamefont {U.}~\bibnamefont {Nowak}},
  \bibinfo {author} {\bibfnamefont {R.}~\bibnamefont {Smirnov-Rueda}}, \bibinfo
  {author} {\bibfnamefont {M.~A.}\ \bibnamefont {Wongsam}}, \bibinfo {author}
  {\bibfnamefont {R.~W.}\ \bibnamefont {Chantrell}},\ and\ \bibinfo {author}
  {\bibfnamefont {J.~M.}\ \bibnamefont {Gonzalez}},\ }\bibfield  {title}
  {\bibinfo {title} {Monte carlo technique with a quantified time step:
  Application to the motion of magnetic moments},\ }\href
  {https://doi.org/10.1103/PhysRevB.67.064422} {\bibfield  {journal} {\bibinfo
  {journal} {Phys. Rev. B}\ }\textbf {\bibinfo {volume} {67}},\ \bibinfo
  {pages} {064422} (\bibinfo {year} {2003})}\BibitemShut {NoStop}%
\bibitem [{\citenamefont {Nowak}\ \emph {et~al.}(2000)\citenamefont {Nowak},
  \citenamefont {Chantrell},\ and\ \citenamefont {Kennedy}}]{TQMC2}%
  \BibitemOpen
  \bibfield  {author} {\bibinfo {author} {\bibfnamefont {U.}~\bibnamefont
  {Nowak}}, \bibinfo {author} {\bibfnamefont {R.~W.}\ \bibnamefont
  {Chantrell}},\ and\ \bibinfo {author} {\bibfnamefont {E.~C.}\ \bibnamefont
  {Kennedy}},\ }\bibfield  {title} {\bibinfo {title} {Monte carlo simulation
  with time step quantification in terms of langevin dynamics},\ }\href
  {https://doi.org/10.1103/PhysRevLett.84.163} {\bibfield  {journal} {\bibinfo
  {journal} {Phys. Rev. Lett.}\ }\textbf {\bibinfo {volume} {84}},\ \bibinfo
  {pages} {163} (\bibinfo {year} {2000})}\BibitemShut {NoStop}%
\bibitem [{\citenamefont {Lent}\ and\ \citenamefont {Kirkner}(1990)}]{QTBM}%
  \BibitemOpen
  \bibfield  {author} {\bibinfo {author} {\bibfnamefont {C.~S.}\ \bibnamefont
  {Lent}}\ and\ \bibinfo {author} {\bibfnamefont {D.~J.}\ \bibnamefont
  {Kirkner}},\ }\bibfield  {title} {\bibinfo {title} {The quantum transmitting
  boundary method},\ }\href {https://doi.org/10.1063/1.345156} {\bibfield
  {journal} {\bibinfo  {journal} {Journal of Applied Physics}\ }\textbf
  {\bibinfo {volume} {67}},\ \bibinfo {pages} {6353} (\bibinfo {year}
  {1990})}\BibitemShut {NoStop}%
\bibitem [{\citenamefont {Tiwari}\ \emph {et~al.}(2019)\citenamefont {Tiwari},
  \citenamefont {de~Put}, \citenamefont {Sor{\'{e}}e},\ and\ \citenamefont
  {Vandenberghe}}]{Tiwari_2019}%
  \BibitemOpen
  \bibfield  {author} {\bibinfo {author} {\bibfnamefont {S.}~\bibnamefont
  {Tiwari}}, \bibinfo {author} {\bibfnamefont {M.~L.~V.}\ \bibnamefont
  {de~Put}}, \bibinfo {author} {\bibfnamefont {B.}~\bibnamefont
  {Sor{\'{e}}e}},\ and\ \bibinfo {author} {\bibfnamefont {W.~G.}\ \bibnamefont
  {Vandenberghe}},\ }\bibfield  {title} {\bibinfo {title} {Carrier transport in
  two-dimensional topological insulator nanoribbons in the presence of vacancy
  defects},\ }\href {https://doi.org/10.1088/2053-1583/ab0058} {\bibfield
  {journal} {\bibinfo  {journal} {2D Materials}\ }\textbf {\bibinfo {volume}
  {6}},\ \bibinfo {pages} {025011} (\bibinfo {year} {2019})}\BibitemShut
  {NoStop}%
\bibitem [{\citenamefont {Metropolis}\ and\ \citenamefont
  {Ulam}(1949)}]{sample9}%
  \BibitemOpen
  \bibfield  {author} {\bibinfo {author} {\bibfnamefont {N.}~\bibnamefont
  {Metropolis}}\ and\ \bibinfo {author} {\bibfnamefont {S.}~\bibnamefont
  {Ulam}},\ }\bibfield  {title} {\bibinfo {title} {The monte carlo method},\
  }\href {https://doi.org/10.1080/01621459.1949.10483310} {\bibfield  {journal}
  {\bibinfo  {journal} {Journal of the American Statistical Association}\
  }\textbf {\bibinfo {volume} {44}},\ \bibinfo {pages} {335} (\bibinfo {year}
  {1949})},\ \bibinfo {note} {pMID: 18139350}\BibitemShut {NoStop}%
\bibitem [{\citenamefont {Akram}\ \emph {et~al.}(2021)\citenamefont {Akram},
  \citenamefont {LaBollita}, \citenamefont {Dey}, \citenamefont {Kapeghian},
  \citenamefont {Erten},\ and\ \citenamefont {Botana}}]{Cr-magnetization}%
  \BibitemOpen
  \bibfield  {author} {\bibinfo {author} {\bibfnamefont {M.}~\bibnamefont
  {Akram}}, \bibinfo {author} {\bibfnamefont {H.}~\bibnamefont {LaBollita}},
  \bibinfo {author} {\bibfnamefont {D.}~\bibnamefont {Dey}}, \bibinfo {author}
  {\bibfnamefont {J.}~\bibnamefont {Kapeghian}}, \bibinfo {author}
  {\bibfnamefont {O.}~\bibnamefont {Erten}},\ and\ \bibinfo {author}
  {\bibfnamefont {A.~S.}\ \bibnamefont {Botana}},\ }\bibfield  {title}
  {\bibinfo {title} {Moire skyrmions and chiral magnetic phases in twisted
  {$\rm CrX_3$} {(X = I, Br, and Cl)} bilayers},\ }\href
  {https://doi.org/10.1021/acs.nanolett.1c02096} {\bibfield  {journal}
  {\bibinfo  {journal} {Nano Letters}\ }\textbf {\bibinfo {volume} {21}},\
  \bibinfo {pages} {6633} (\bibinfo {year} {2021})},\ \bibinfo {note} {pMID:
  34339218}\BibitemShut {NoStop}%
\bibitem [{\citenamefont {Kapoor}\ \emph {et~al.}(2021)\citenamefont {Kapoor},
  \citenamefont {Mandal}, \citenamefont {Adak}, \citenamefont {Patankar},
  \citenamefont {Manni}, \citenamefont {Thamizhavel},\ and\ \citenamefont
  {Deshmukh}}]{alphaCrCl3}%
  \BibitemOpen
  \bibfield  {author} {\bibinfo {author} {\bibfnamefont {L.~N.}\ \bibnamefont
  {Kapoor}}, \bibinfo {author} {\bibfnamefont {S.}~\bibnamefont {Mandal}},
  \bibinfo {author} {\bibfnamefont {P.~C.}\ \bibnamefont {Adak}}, \bibinfo
  {author} {\bibfnamefont {M.}~\bibnamefont {Patankar}}, \bibinfo {author}
  {\bibfnamefont {S.}~\bibnamefont {Manni}}, \bibinfo {author} {\bibfnamefont
  {A.}~\bibnamefont {Thamizhavel}},\ and\ \bibinfo {author} {\bibfnamefont
  {M.~M.}\ \bibnamefont {Deshmukh}},\ }\bibfield  {title} {\bibinfo {title}
  {Observation of standing spin waves in a van der waals magnetic material},\
  }\href {https://doi.org/https://doi.org/10.1002/adma.202005105} {\bibfield
  {journal} {\bibinfo  {journal} {Advanced Materials}\ }\textbf {\bibinfo
  {volume} {33}},\ \bibinfo {pages} {2005105} (\bibinfo {year}
  {2021})}\BibitemShut {NoStop}%
\bibitem [{\citenamefont {Zhong}\ \emph
  {et~al.}(2017{\natexlab{b}})\citenamefont {Zhong}, \citenamefont {Seyler},
  \citenamefont {Linpeng}, \citenamefont {Cheng}, \citenamefont {Sivadas},
  \citenamefont {Huang}, \citenamefont {Schmidgall}, \citenamefont {Taniguchi},
  \citenamefont {Watanabe}, \citenamefont {McGuire}, \citenamefont {Yao},
  \citenamefont {Xiao}, \citenamefont {Fu},\ and\ \citenamefont
  {Xu}}]{FM-hetero}%
  \BibitemOpen
  \bibfield  {author} {\bibinfo {author} {\bibfnamefont {D.}~\bibnamefont
  {Zhong}}, \bibinfo {author} {\bibfnamefont {K.~L.}\ \bibnamefont {Seyler}},
  \bibinfo {author} {\bibfnamefont {X.}~\bibnamefont {Linpeng}}, \bibinfo
  {author} {\bibfnamefont {R.}~\bibnamefont {Cheng}}, \bibinfo {author}
  {\bibfnamefont {N.}~\bibnamefont {Sivadas}}, \bibinfo {author} {\bibfnamefont
  {B.}~\bibnamefont {Huang}}, \bibinfo {author} {\bibfnamefont
  {E.}~\bibnamefont {Schmidgall}}, \bibinfo {author} {\bibfnamefont
  {T.}~\bibnamefont {Taniguchi}}, \bibinfo {author} {\bibfnamefont
  {K.}~\bibnamefont {Watanabe}}, \bibinfo {author} {\bibfnamefont {M.~A.}\
  \bibnamefont {McGuire}}, \bibinfo {author} {\bibfnamefont {W.}~\bibnamefont
  {Yao}}, \bibinfo {author} {\bibfnamefont {D.}~\bibnamefont {Xiao}}, \bibinfo
  {author} {\bibfnamefont {K.-M.~C.}\ \bibnamefont {Fu}},\ and\ \bibinfo
  {author} {\bibfnamefont {X.}~\bibnamefont {Xu}},\ }\bibfield  {title}
  {\bibinfo {title} {Van der waals engineering of ferromagnetic semiconductor
  heterostructures for spin and valleytronics},\ }\href
  {https://doi.org/10.1126/sciadv.1603113} {\bibfield  {journal} {\bibinfo
  {journal} {Science Advances}\ }\textbf {\bibinfo {volume} {3}},\ \bibinfo
  {pages} {e1603113} (\bibinfo {year} {2017}{\natexlab{b}})}\BibitemShut
  {NoStop}%
\end{thebibliography}%
\end{document}